\documentclass[aps,prd,twocolumn,showpacs,superscriptaddress,groupedaddress]{revtex4-2}  
\usepackage{graphicx}  
\usepackage{dcolumn}   
\usepackage{bm}        
\usepackage{amssymb}   

\usepackage{lineno}
\usepackage{booktabs}
\usepackage{longtable}
\usepackage{overpic}
\usepackage{multirow}
\usepackage{amsmath}
\usepackage{color}
\usepackage[bookmarksnumbered, pdfstartview=FitH,colorlinks,urlcolor=blue, citecolor=blue,linkcolor=blue]{hyperref}

\usepackage{subfigure}
\usepackage{tabularx}
\usepackage{textcomp}
\usepackage{gensymb}
\usepackage{cleveref}

\crefname{section}{Sec.}{Secs.}
\crefname{figure}{Fig.}{Figs.}
\begin{document}



\title{\boldmath Measurement of
  \texorpdfstring{$e^+e^-\rightarrow\Lambda\bar{\Lambda}\eta$}{eetoLambdaLambdabarEta} from 3.5106 to 4.6988 GeV and study of
  \texorpdfstring{$\Lambda\bar{\Lambda}$}{LambdaLambdabar}
mass threshold enhancement}

\author{
\begin{small}
\begin{center}
M.~Ablikim$^{1}$, M.~N.~Achasov$^{12,b}$, P.~Adlarson$^{72}$, M.~Albrecht$^{4}$, R.~Aliberti$^{33}$, A.~Amoroso$^{71A,71C}$, M.~R.~An$^{37}$, Q.~An$^{68,55}$, Y.~Bai$^{54}$, O.~Bakina$^{34}$
% ...
\\
\vspace{0.2cm}
(BESIII Collaboration)\\
\vspace{0.2cm} {\it
$^{1}$ Institute of High Energy Physics, Beijing 100049, People's Republic of China\\
% ...
$^{4}$ Bochum Ruhr-University, D-44780 Bochum, Germany\\
% ...
$^{12}$ G.I. Budker Institute of Nuclear Physics SB RAS (BINP), Novosibirsk 630090, Russia\\
% ...
$^{33}$ Johannes Gutenberg University of Mainz, Johann-Joachim-Becher-Weg 45, D-55099 Mainz, Germany\\
$^{34}$ Joint Institute for Nuclear Research, 141980 Dubna, Moscow region, Russia\\
% ...
$^{37}$ Liaoning Normal University, Dalian 116029, People's Republic of China\\
% ...
$^{54}$ Southeast University, Nanjing 211100, People's Republic of China\\
$^{55}$ State Key Laboratory of Particle Detection and Electronics, Beijing 100049, Hefei 230026, People's Republic of China\\
% ...
$^{68}$ University of Science and Technology of China, Hefei 230026, People's Republic of China\\
% ...
$^{71}$ University of Turin and INFN, (A)University of Turin, I-10125, Turin, Italy; (B)University of Eastern Piedmont, I-15121, Alessandria, Italy; (C)INFN, I-10125, Turin, Italy\\
$^{72}$ Uppsala University, Box 516, SE-75120 Uppsala, Sweden\\
% ...
$^{b}$ Also at the Novosibirsk State University, Novosibirsk, 630090, Russia\\
% ...
}\end{center}

\vspace{0.4cm}
\end{small}

}

\begin{abstract}
Using data samples with a total integrated luminosity of
approximately 18 fb$^{-1}$ collected by the BESIII detector operating
at the BEPCII, the process $e^+e^-\rightarrow\Lambda\bar{\Lambda}
\eta$ is studied at center-of-mass energies between 3.5106 and 4.6988
GeV. The Born cross section for the process
$e^+e^-\rightarrow\Lambda\bar{\Lambda}\eta$ is measured. No
significant structure is observed in the Born cross section line
shape. An enhancement near the $\Lambda\bar{\Lambda}$ mass
threshold is observed for the first time in the process. The structure can be
described by an $S$-wave Breit-Wigner function. Neglecting
contribution of excited $\Lambda$ states and potential interferences, the mass and width are determined to be ($2356\pm
7\pm17$) MeV/$c^2$ and ($304\pm28\pm54$) MeV, respectively,
where the first uncertainties are statistical and the second are
systematic.
\end{abstract}

\pacs{13.60.Rj, 13.66.De, 13.85.Lg, 14.40.Rt}

\maketitle

\section{Introduction}

In 2005, the BABAR collaboration reported a structure, the $Y(4260)$,
with a mass around 4.26 GeV/$c^2$  in the $\pi^+\pi^-J/\psi$ final state via
the initial state radiation~(ISR) process $e^{+}e^{-}
\rightarrow\gamma_{\mathrm{ISR}}\pi^{+}\pi^{-}
J/\psi$~\cite{ObservationY4260_PhysRevLett}. The observation was
confirmed by the CLEO~\cite{CleoY4260_PhysRevD.74.091104} and
Belle~\cite{BelleY4260_PhysRevLett} collaborations. Given its strong
coupling into the charmonium state, it must contain a
charm-anticharm quark pair, because of the strong suppression of the
heavy quark-antiquark pair creation within Quantum
Chromodynamics~\cite{Brambilla:2019esw}. Because the $Y(4260)$ can be
produced by a virtual photon from $e^+e^-$ annihilation, it is a
vector state with the spin parity $J^{PC}$ of $1^{--}$. Above the
$D\bar{D}$ threshold, the four vector charmonium states~[$\psi(3770)$,
  $\psi(4040)$, $\psi(4160)$, and $\psi(4415)$] predicted by the
potential model~\cite{Voloshin:2007dx_pm} have been
established~\cite{ParticleDataGroup:2020ssz}. Moreover, the
traditional charmonium states dominantly couple to the ground-state
open-charm meson pairs instead of hidden charm
states~\cite{ParticleDataGroup:2020ssz}.  Therefore, the $Y(4260)$
cannot be explained as a traditional charmonium state. A more precise
measurement performed by BESIII collaboration shows the $Y(4260)$
contains two structure $Y(4220)$ and $Y(4330)$~\cite{BESIII:2016bnd}.
Later, more
states with similar properties were
observed~\cite{BaBar:2006ait,Belle:2007umv}. These $Y$ states, named
$\psi$ states in the
Particle Data Group~(PDG)~\cite{ParticleDataGroup:2020ssz}, are good
candidates for exotic states, such as glueballs, compact tetraquarks,
hybrids, hadrocharmonia, etc.~\cite{Brambilla:2019esw}. Some of these
interpretations, however, are disfavored by experiment~\cite{Zhu:2005hp}.  Light
hadron final states of $Y$ decays, such as $\pi^0p\overline{p}
$~\cite{pi0pp_201745}, $ K^0_{S}K^{\pm}\pi^{\mp}
$~\cite{KKpi_PhysRevD.99.072005}, $K_{S}^{0}
K^{\pm} \pi^{\mp} \pi^{0}(\eta)$~\cite{BESIII:2018kyw}, $\eta \phi \pi^{+} \pi^{-}$~\cite{BESIII:2017qkh} $ 
\Xi^{-}\bar{\Xi}^{+} $~\cite{eeXiXi_PhysRevLett.124.032002},
$2(p\overline{p})$~\cite{BESIII:2020svk},
$\Lambda\bar\Lambda$~\cite{BESIII:2021ccp}, and $\omega\pi^{0}$~\cite{BESIII:2022zxr} have been
searched for but not yet discovered. Searching for new
decay modes of $Y$ states, for instance $Y \to
\Lambda\bar{\Lambda}\eta$, can add more information to our
understanding of their inner structure.

Enhancements near the baryon-antibaryon mass threshold have been
observed in low energy $p\overline{p}$ collisions, charmonium decays,
and $B$ meson decays. Using PS185 low energy $p\overline{p}$ collision
data, a structure near the $\Lambda\bar{\Lambda}$ threshold was
observed with a partial wave analysis of $p\overline{p} \to\Lambda
\bar{\Lambda}$~\cite{Bugg:2004rj}.  In charmonium decays, enhancements
are observed in $p\overline{p}$ final states of the processes
$J/\psi\rightarrow \gamma
p\overline{p}$~\cite{Jpsippgamma_PhysRevLett.91.022001}.
In
$B$ meson decays, enhancements were found in the processes
$B^0\rightarrow\Lambda\bar{\Lambda}K^0(K^{*0})$~\cite{Chang:2008yw}, $B^{0} \rightarrow p \bar{\Lambda}
\pi^{-}$~\cite{BaBar:2009ess}, $B^{+} \rightarrow p \overline{p}
\pi^{+}, B^{0} \rightarrow p \overline{p}
K^{0}$~\cite{Belle:2007oni}, $B^{+} \rightarrow \Lambda \bar{\Lambda}
K^{+}$~\cite{Belle:2008kpj}, $B^{+(0)} \rightarrow p \bar{p} K^{*+(0)}$~\cite{Belle:2008zkc}, $B^{0} \rightarrow \bar{D}^{* 0} p
\overline{p}, \bar{D}^{0} p \overline{p}$~\cite{BaBar:2011zka}, $B^{0}
\rightarrow p\bar{\Lambda} D^{(*)-}$~\cite{Belle:2015crh}, $B^0_{(s)} \to p
\bar{p} h^+ h^{\prime-}$~\cite{LHCb:2017obv}, $B^{+}
\rightarrow p\bar{\Lambda} K^+ K^-$~\cite{Belle:2018ies}, $B^0\to
p\bar{p}\pi^0$~\cite{Belle:2019abe}, $B^{+}\to
p\overline{p}\mu^{+}\nu_{\mu}$~\cite{LHCb:2019cgl}
etc.  Strikingly, there is no corresponding enhancement observed in
$\psi(2S)\rightarrow\gamma p \overline{p}$~\cite{CLEO:2010fre}, $\psi(2S)
\rightarrow p\bar{p}\pi^0$~\cite{BESIII:2012ssm},
$\Upsilon(1S)\to\gamma p \overline{p}$~\cite{CLEO:2005koa},
$J/\psi\to\omega p \overline{p}$~\cite{BESIII:2013lac}, $\psi(2 S) \rightarrow
p \overline{p} \eta$~\cite{BESIII:2013xkm}, $J/\psi\text{,~}\psi(2S)\to
p\overline{p}\phi$~\cite{BESIII:2015dag,BESIII:2019rth},
$J/\psi\text{,~}\psi(2S)\rightarrow\eta \Sigma^{+}
\bar{\Sigma}^{-}$~\cite{BESIII:2022ahw}, $\psi(2S) \rightarrow \Lambda
\bar{\Lambda} \eta~(\pi^0)$~\cite{BESIII:2022cxi},
$\psi(2S)\rightarrow\Lambda\bar{\Lambda}\omega$~\cite{BESIII:2022fhe},
$B^{+} \to p \bar p K^{+}$~\cite{LHCb:2013ala}, $\bar{B}^0 \to
D^0\Lambda\bar{\Lambda}$~\cite{BaBar:2014cco}, $B
\to p {\bar p} \pi \pi$~\cite{Belle:2019fnj}
$B_s^0 \to J/\psi p \bar{p}$~\cite{LHCb:2021chn}, and $B^-\to J/\psi\Lambda
\overline{p}$~\cite{LHCb:2022jad} decays. Many theory models have
been proposed for the interpretation of these enhancements, including
a gluonic mechanism~\cite{ThresBdecay_Rosner:2003bm}, a fragmentation
mechanism~\cite{ThresBdecay_Rosner:2003bm},
baryonia~\cite{Deng:2013aca}, multiquark states~\cite{Ding:2005tr},
and final state interactions~\cite{Dalkarov:2009yf}. Searching for
enhancements with different quantum numbers in different reactions can
provide more constraints on these models.

In this article, we report a study of the reaction
$e^+e^-\rightarrow\Lambda\bar{\Lambda}\eta$ with a partial
reconstruction technique, based on the data recorded at center-of-mass
energies from 3.5106 to 4.6988 GeV. The Born cross sections are
measured, and possible resonances are searched for by fitting the
energy-dependent distribution of the Born
cross sections. In addition, a threshold enhancement is observed in the
$\Lambda\bar{\Lambda}$ mass spectrum. The mass and width of the
structure are determined by a fit with an $S$-wave Breit-Wigner
function, and the angular distribution of the structure is studied.

\section{Detector and data samples}
\label{sec:detSample}
The BESIII detector~\cite{Ablikim:2009aa} records symmetric $e^+e^-$ collisions 
provided by the BEPCII storage ring~\cite{Yu:2016cof}, which operates 
in the center-of-mass energy range from 2.00 to 4.95~GeV.
The cylindrical core of the BESIII detector covers 93\% of the full solid angle and consists of a helium-based
 multilayer drift chamber~(MDC), a plastic scintillator time-of-flight
system~(TOF), and a CsI(Tl) electromagnetic calorimeter~(EMC),
which are all enclosed in a superconducting solenoidal magnet
providing a 1.0~T 
magnetic field. The solenoid is supported by an
octagonal flux-return yoke with resistive plate counter muon
identification modules interleaved with steel. 
The charged-particle momentum resolution at $1~{\rm GeV}/c$ is
$0.5\%$, and the ${\rm d}E/{\rm d}x$ resolution is $6\%$ for electrons
from Bhabha scattering. The EMC measures photon energies with a
resolution of $2.5\%$ ($5\%$) at $1$~GeV in the barrel (end cap)
region. The time resolution in the TOF barrel region is 68~ps, while
that in the end cap region is 110~ps. The end cap TOF
system was upgraded in 2015 using multi-gap resistive plate chamber
technology, providing a time resolution of
60~ps~\cite{Cao:2020ibk}.

Simulated data samples produced with {\sc
  geant4}-based~\cite{GEANT4:2002zbu} Monte Carlo (MC) software, which
includes the geometric description of the BESIII detector and the
detector response, are used to determine detection efficiencies and to
estimate background contributions. The simulation includes the beam
energy spread and ISR in the $e^+e^-$ annihilations modeled with the
generator {\sc kkmc}~\cite{Jadach:2000irKKMC, *Jadach:1999vf}. The
inclusive MC sample includes the production of open charm processes,
the ISR production of vector charmonium(-like) states, and the
continuum processes incorporated in {\sc
  kkmc}~\cite{Jadach:2000irKKMC, *Jadach:1999vf}. All particle decays
are modeled with {\sc evtgen}~\cite{Lange:2001uf,
  *Ping:2008zz_BesGen} using branching fractions either taken from
  the PDG~\cite{ParticleDataGroup:2020ssz}, when
available, or otherwise estimated with {\sc
  lundcharm}~\cite{Chen:2000tv, *Yang:2014vra}. Final state radiation
from charged final state particles is incorporated using {\sc
  photos}~\cite{Richter-Was:1992hxq}. At each energy point, we
generate a signal MC sample of three-body process
$e^+e^-\rightarrow\Lambda\bar{\Lambda}\eta$ with
$\eta\rightarrow\gamma\gamma$ using a uniformly distributed phase
space~(PHSP) model. The MC sample for the study of threshold
enhancement is discussed in Sec~\ref{sec:StudyofMLambdabarLambda}.

\section{Event Selection}
\label{:sec:evetSelect}

The signal candidates of the process
$e^+e^-\rightarrow\Lambda\bar{\Lambda}\eta$ are selected with a
partial reconstruction technique, to achieve higher efficiency. This
technique requires that only the $\Lambda$~($\bar{\Lambda}$) baryon
and $\eta$ meson are reconstructed, inferring the
$\bar{\Lambda}$~($\Lambda$) from energy-momentum conservation. The
$\Lambda$~($\bar{\Lambda}$) candidates and the $\eta$ candidates are
reconstructed with $p\pi^-$~($\overline{p}\pi^{+}$) and $\gamma\gamma$
decay modes, respectively.

Charged tracks detected in the MDC are required to be within a polar
angle ($\theta$) range of $|\!\cos\theta|<0.93$, where $\theta$ is
defined with respect to the $z$-axis, the symmetry axis of
the MDC. The distance of the closest approach to the interaction point
(IP) along the $z$-axis must be less than 20\,cm. Particle
identification~(PID) for charged tracks combines measurements of ${\rm
  d}E/{\rm d}x$ in the MDC and the flight time in the TOF to form
likelihoods for each hadron~(pion, kaon, and proton)
hypothesis. Tracks are identified as protons~(pions) when the
proton~(pion) hypothesis has the largest likelihood. Proton
and pion track pairs with opposite charges are constrained to originate
from a common vertex to form the $\Lambda$~($\bar{\Lambda}$)
candidates by a vertex fit~\cite{Xu:2009zzg}. The obtained $\chi^2$ in the
fit, denoted as $\chi^2_\text{vtx}$,
are kept for further analysis. The decay length of the
$\Lambda$~($\bar{\Lambda}$) candidate is required to be greater than twice the
vertex resolution away from the IP\@.

Photon candidates are identified using showers in the EMC.  The
deposited energy of each shower must be more than 25~MeV in the barrel
region ($|\!\cos\theta|< 0.80$) and more than 50~MeV in the end cap
region ($0.86 <|\!\cos\theta|< 0.92$). To exclude showers that
originate from charged tracks identified as antiprotons, the angle
subtended by the EMC shower and the position of the closest charged
track at the EMC must be greater than 20 degrees as measured from the
IP. The angle is required to be greater than 10 degrees for other
types of charged tracks. To suppress electronic noise and showers
unrelated to the event, the difference between the EMC time and the
event start time is required to be within [0, 700]\,ns. To remove fake
photons from neutron-antineutron annihilations, $E_{3 \times 3} /
E_{5 \times 5}$ and the lateral moment~\cite{LateralMM_Drescher:1984rt}
describing the shape of electromagnetic showers are required to be less than
$0.4$ and greater than $0.85$, respectively, where $E_{3 \times
  3}(E_{5 \times 5})$ is total energy deposited in the $3 \times
3$~($5 \times 5$) crystals around the seed of the
shower~\cite{He:2011zzd}. Photon pairs are used to reconstruct
$\eta\to \gamma\gamma$ decays.

To suppress the background and improve the momentum resolution, a
one-constraint~(1C) kinematic fit~\cite{asner2008physics} under the
hypothesis $e^+e^- \to \gamma\gamma p\pi^-\overline{p}\pi^+$ is
performed with the recoil mass against $\gamma\gamma
p\pi^-$~($\gamma\gamma \overline{p} \pi^+$) constrained to the known
mass of $\Lambda$, $M_{\Lambda}$~\cite{ParticleDataGroup:2020ssz}. In
case of multiple reconstructed $\gamma\gamma p\pi^-$~($\gamma\gamma
\overline{p} \pi^+$) candidates, the candidate with the smallest
$\chi^2_\text{vtx}+\chi^2_\text{1C}$ is kept, where
$\chi^2_\text{vtx}$ and $\chi^2_\text{1C}$ are the $\chi^2$ of the
vertex fit and 1C kinematic fit, respectively. We require 
$\chi^2_\text{vtx}<20$ and the $\chi^2_\text{1C}<10$.  After the above selection criteria are applied, the
invariant mass of the $p\pi^-$~($\overline{p}\pi^+$) combination is
required to satisfy $1.112 \text{~GeV}/c^2
<M_{p\pi^-~(\overline{p}\pi^+)}<1.120 \text{~GeV}/c^2$. A study with
the inclusive MC sample using TopoAna~\cite{Zhou:2020ksj} shows that
the main background contributions are from processes with
$\Sigma^0$~($\bar{\Sigma}^0$) baryons in the final states. The
$\Sigma^0$~($\bar{\Sigma}^0$) baryon decays to
$\Lambda\gamma$~($\bar{\Lambda}\gamma$) with a branching fraction of
100\%~\cite{ParticleDataGroup:2020ssz}.  In order to veto
processes with a soft $\gamma_L$ from $\Sigma^0\to
\Lambda\gamma_L$~($\bar{\Sigma}^0\to \bar{\Lambda}\gamma_L$) decays,
the invariant mass of the reconstructed $\Lambda$~($\bar{\Lambda}$)
and the selected photon with lower energy
$M_{\Lambda\gamma_{L}}$~($M_{\bar{\Lambda}\gamma_{L}}$) is required to
be outside the $\Sigma^0$~($\bar{\Sigma}^0$) mass region of [1.113,
  1.273] GeV/$c^2$.
A study of the $M_{\gamma\gamma}$ distribution with the inclusive MC
indicates small peaking background contributions from the processes
$e^+e^-\rightarrow\eta\Sigma^{0} \bar{\Lambda}+\text{c}.\text{c}$ and
$e^+e^-\rightarrow\Sigma^0\bar{\Sigma}^0\eta$ near the $\eta$ peak in
the $M_{\gamma\gamma}$ distribution. However, the process
$e^+e^-\rightarrow\eta\Sigma^{0}\bar{\Lambda}+\text{c}.\text{c}$ is
suppressed because of isospin violation. Due to the two extra photons
in the final state, the $e^+e^-\rightarrow\Sigma^0\bar{\Sigma}^0\eta$
process is also suppressed by the 1C kinematic fit.

\section{Cross Section Measurement}%
\label{sec:CS}
 
The Born cross section of the process
$e^+e^-\rightarrow\Lambda\bar{\Lambda}\eta$ with contributions from
intermediate states at each center-of-mass
energy is calculated as

\begin{equation}
  \label{eq:csborn}
  \begin{aligned}
    \sigma^{\mathrm{Born}}=\frac{N_{\mathrm{sig}}}{\mathcal{L}_\mathrm{{int}}\mathcal{B}\varepsilon(1+\delta)(1+\delta^{v})},
  \end{aligned}
\end{equation}
where $N_{\mathrm{sig}}$ is the signal yield,
$\mathcal{L}_{\mathrm{int}}$ is the integrated luminosity measured
using large-angle Bhabha events~\cite{BESIII:2015qfd, BESIII:2022dxl,
  BESIII:2022Lum}, $\mathcal{B}$ is the product of the branching
fractions of $\eta\rightarrow\gamma\gamma$ and $\Lambda\rightarrow
p\pi^-$~($\bar{\Lambda}\rightarrow \overline{p}\pi^+$) decays
from the PDG~\cite{ParticleDataGroup:2020ssz}, $\varepsilon$ is the
reconstruction efficiency determined by MC simulation, $(1+\delta)
$ is the ISR correction factor, and $(1+\delta^{v})$ is the vacuum
polarization factor~\cite{WorkingGrouponRadiativeCorrections:2010bjp}.

\begin{figure}[t]
  \centering
  \begin{overpic}[width=0.45\textwidth, percent] {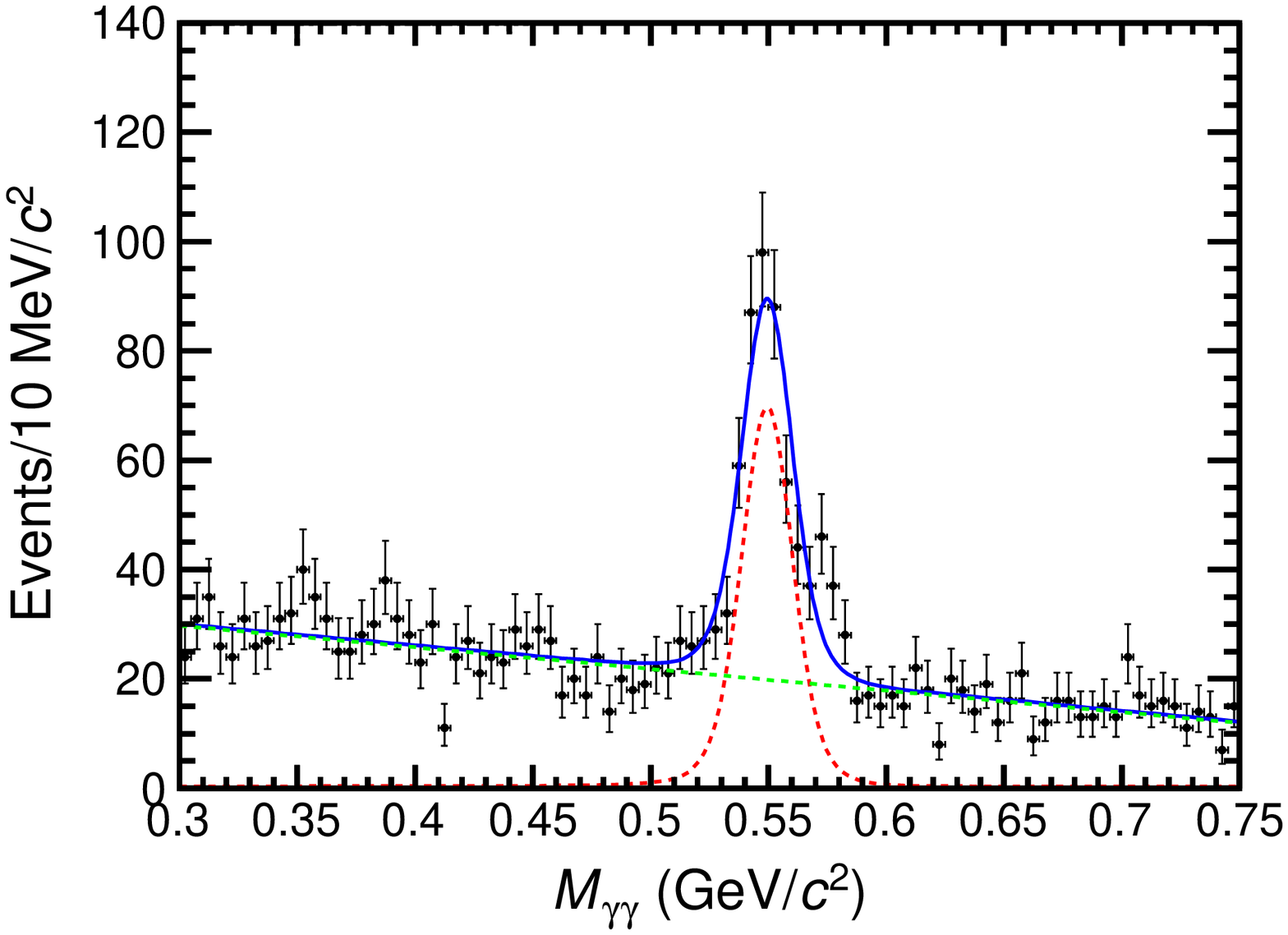}\put(80,60){(b)} \end{overpic}
  \begin{overpic}[width=0.45\textwidth, percent] {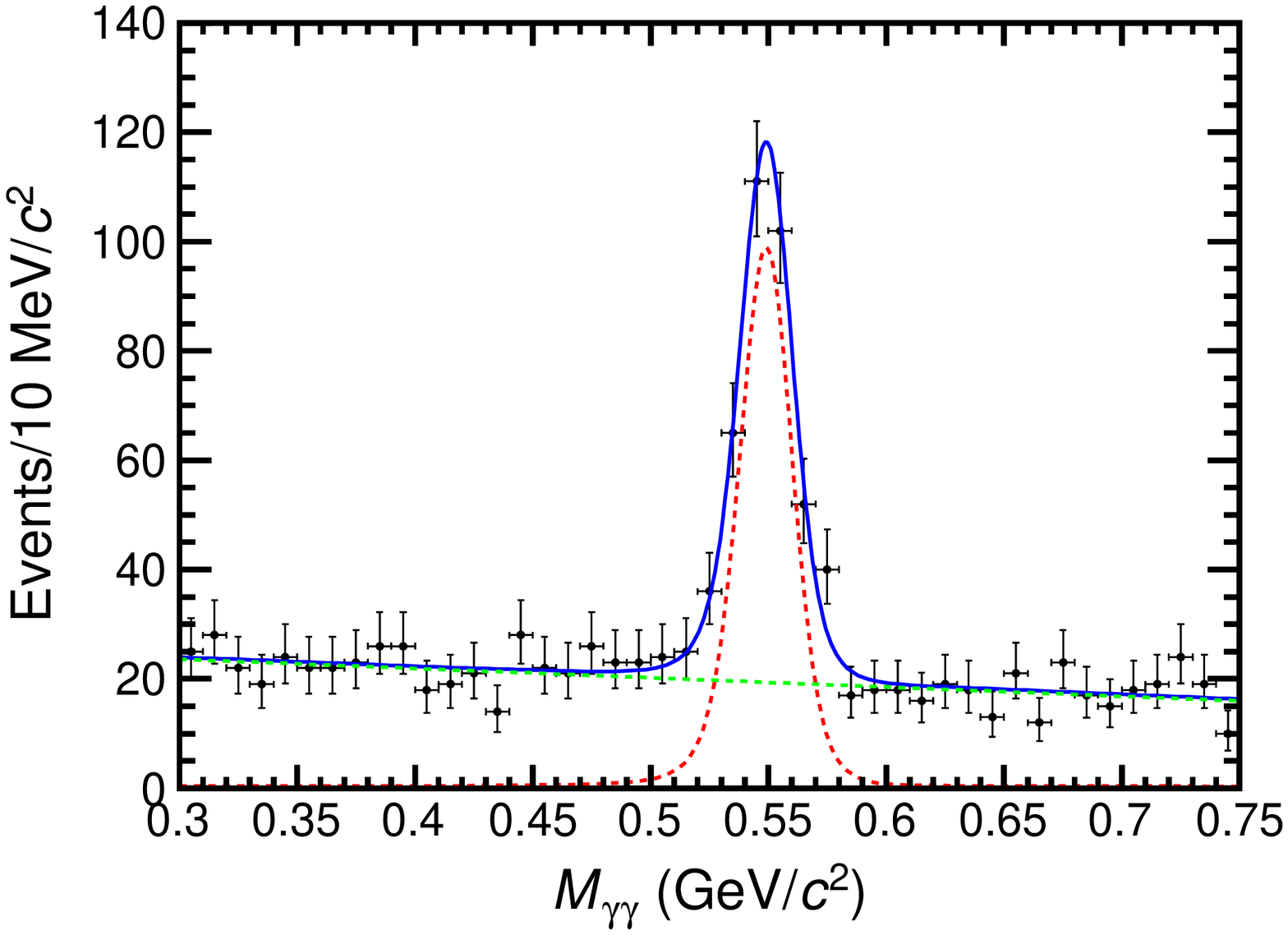}\put(80,60){(b)} \end{overpic}
  \caption{Fits to the invariant mass spectra of $\gamma\gamma$ for
  the events surviving the selection criteria at (a)~$\sqrt{s}=3.7730$ and
  (b)~$\sqrt{s}=4.1784$
GeV. The black points with error bars are data. The solid blue,
dashed red, and dashed green lines correspond to the fit result, signal, and
background, respectively.}\label{fig:massspecGG}
\end{figure}

The signal yield is determined by an unbinned maximum-likelihood fit
to the $M_{\gamma\gamma}$ spectrum at each energy point. The
background is modeled by a linear function. The signal is described by
an MC-simulated shape convolved with a Gaussian function which is used
to compensate for the differences in resolution between the MC
simulation and the data. The means of the Gaussian functions at different
energy points are free. The widths of the Gaussian functions cannot be
constrained by data samples with low luminosities. Therefore, they
are fixed to a common value obtained from the fit of the data sample
with the highest luminosity taken at $\sqrt{s} =$3.7730
GeV. Figure~\ref{fig:massspecGG} shows the fit results for the data at
$\sqrt{s} =$4.1784 GeV. The number of signal events in the $\eta$
signal region~($0.521<M_{\gamma\gamma}<0.572$ GeV/$c^2$) for each
energy point is determined by integrating the fitted signal shape in
the signal region and is summarized in Table~\ref{tab:cswitherr}. The
contribution of peaking background mentioned in
Sec.~\ref{:sec:evetSelect} is estimated to be 1.34\% at $\sqrt{s}
=4.1784\text{~GeV}$ by adding the shape derived from the MC simulation
of the two processes to the nominal fit model. The effect of the
peaking background events on the determination of signal yields will
be discussed in Sec.~\ref{sec:sysErr}.

\begin{figure}[t]
  \centering
  \begin{overpic}[width=0.235\textwidth, percent]{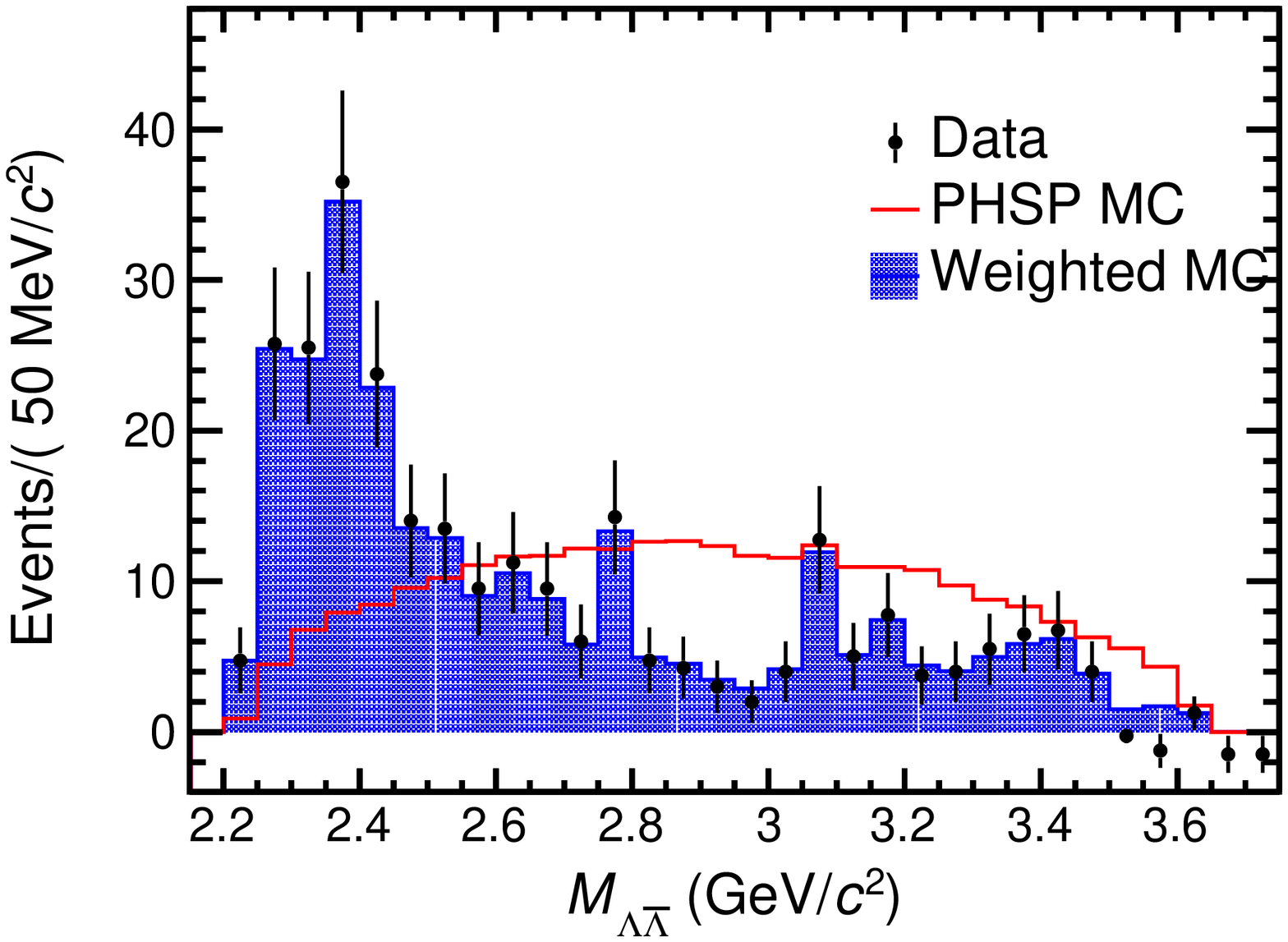}\put(80,60){(a)}
  \end{overpic}
  \begin{overpic}[width=0.235\textwidth, percent]{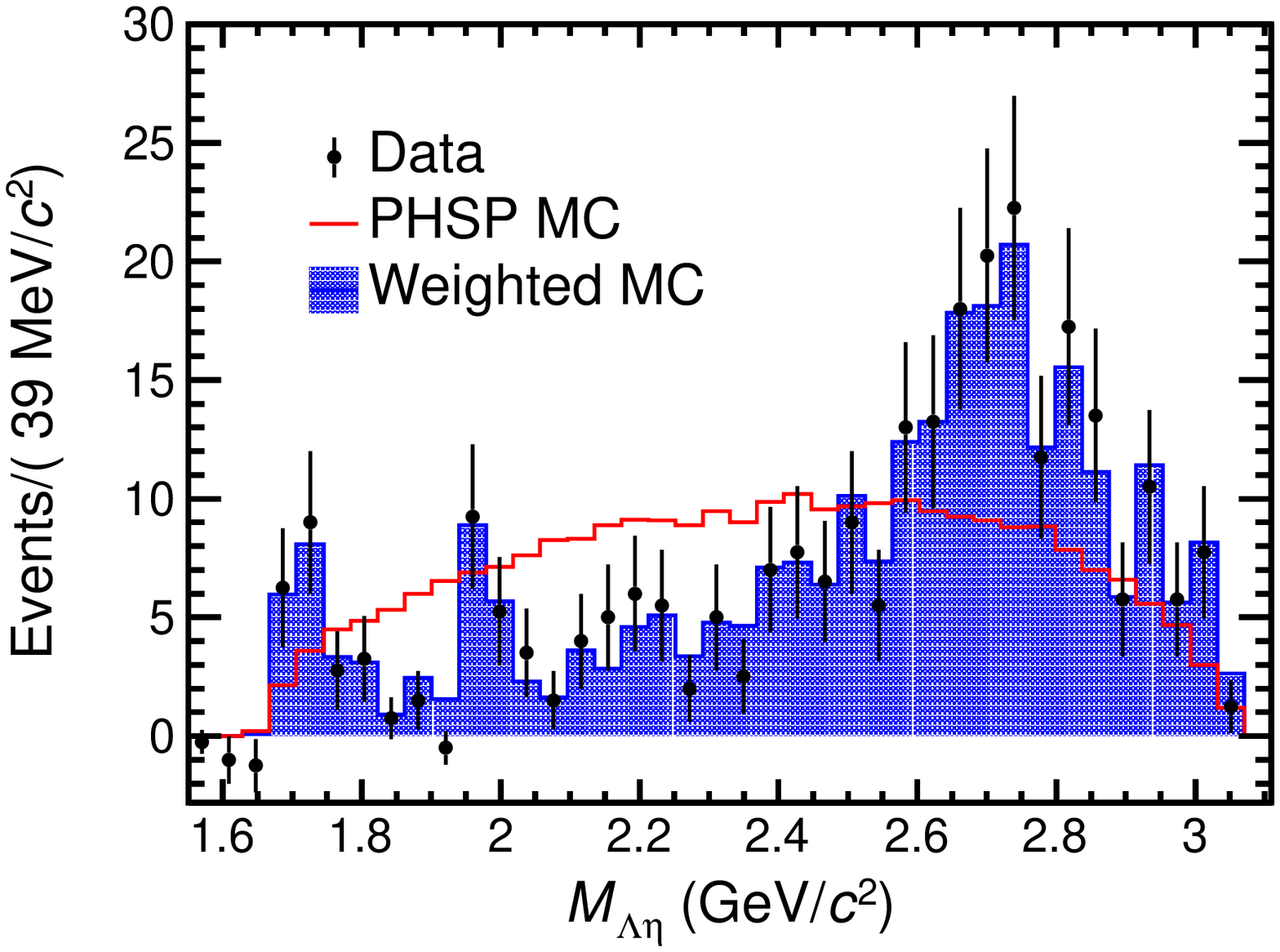}
    \put(80,60){(b)}
  \end{overpic}\\
  \begin{overpic}[width=0.235\textwidth, percent]{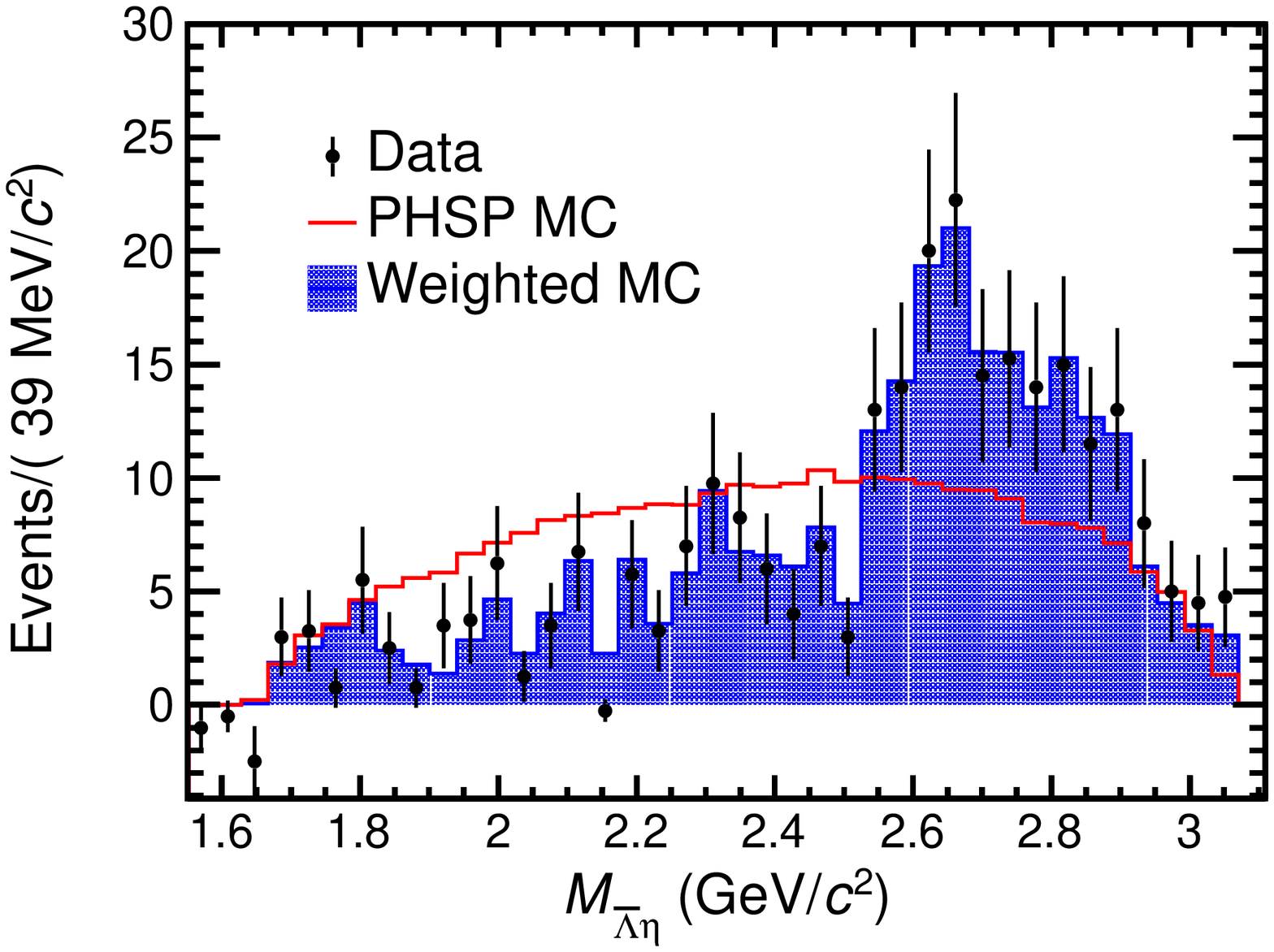}
    \put(80,60){(c)}
  \end{overpic}
  \caption{Distributions of (a) $M_{\Lambda\Bar{\Lambda}}$, (b)
  $M_{\Lambda\eta}$, and (c) $M_{\Bar{\Lambda}\eta}$ obtained from
events within the $\eta$ signal region at $\sqrt{s} =4.1784$ GeV. The
black points with error bars are background-subtracted data. The
solid~(red) histograms and hatched~(blue) histograms denote the PHSP
MC and weighted PHSP MC sample, respectively.}\label{fig:mcdataComp}
\end{figure}

As shown in Fig.~\ref{fig:mcdataComp}, evident discrepancies between
the background-subtracted data and PHSP MC simulation are seen, where
the background contributions are subtracted using the events from the
$\eta$ sidebands, defined as $0.368<M_{\gamma\gamma}< 0.470$ GeV/$c^2$
and $0.622<M_{\gamma\gamma}< 0.724\mathrm{~GeV}/c^2$. To estimate the
reconstruction efficiencies, the PHSP MC samples are weighted
according to the background-subtracted data distributions of
$M_{\Lambda\Bar{\Lambda}}$, $M_{\Lambda\eta}$, and $M_{\Bar{\Lambda}
  \eta}$ shown in Fig.~\ref{fig:mcdataComp}. Because of limited sample
sizes, we can not determine the exact two-dimensional weighting
factors from the $e^+e^-\to\Lambda\bar{\Lambda}\eta$ Dalitz
distributions of data. We take the weight for each two-dimensional
distribution as the product of the one dimensional distributions. The
distributions of $M_{\Lambda\bar{\Lambda}}$, $M_{\Lambda\eta}$, and
$M_{\bar{\Lambda}\eta}$ are treated as independent. Because of this
approximation, all combinations of invariant masses among $\Lambda$,
$\bar{\Lambda}$, and $\eta$ are taken into account to make the MC
simulation consistent with data.  Therefore, the weighting factor
is calculated as the average of the three two-dimensional distributions

\begin{equation}
  \label{eq:MCweightingfactor}
  \begin{aligned}
   & w^{i=\{m,n,o\}} (M_{\Lambda \bar{\Lambda}}, M_{\Lambda \eta}, M_{\bar{\Lambda}
    \eta}) =  \frac{1}{3}  \frac{
    N^{\text{data}}_{M_{\Lambda \bar{\Lambda}}}(m)}{N^{\mathrm{MC}}_{M_{\Lambda
    \bar{\Lambda}}}(m)}  \frac{
    N^{\text{data}}_{M_{\Lambda \eta}}(n)}{N^{\mathrm{MC}}_{M_{\Lambda
          \eta}}(n)} \\
    & +  \frac{1}{3}  \frac{
    N^{\text{data}}_{M_{\Lambda \bar{\Lambda}}}(m)}{N^{\mathrm{MC}}_{M_{\Lambda
    \bar{\Lambda}}}(m)}  \frac{
    N^{\text{data}}_{M_{\bar{\Lambda} \eta}}(o)}{N^{\mathrm{MC}}_{M_{\bar{\Lambda}\eta}}(o)}
    +  \frac{1}{3}  \frac{
    N^{\text{data}}_{M_{\Lambda \eta}}(n)}{N^{\mathrm{MC}}_{M_{\Lambda
    \eta}}(n)}  \frac{
    N^{\text{data}}_{M_{\bar{\Lambda} \eta}}(o)}{N^{\mathrm{MC}}_{M_{\bar{\Lambda}\eta}}(o)},
  \end{aligned}
\end{equation}
where $w^i$ is the weight for event $i$ corresponding to bins
$m$, $n$, and $o$ of the $M_{\Lambda\bar{\Lambda}}$, $M_{\bar{\Lambda}
  \eta}$ and $M_{\bar{\Lambda}\eta}$ distributions,
$N^{\text{data}/\mathrm{MC}}_{M_{\Lambda\bar{\Lambda}}}(m)$,
$N^{\text{data}/\mathrm{MC}}_{M_{\Lambda \eta}}(n)$, and
$N^{\text{data}/\mathrm{MC}}_{M_{\bar\Lambda \eta}}(o)$ are numbers of
events in bins $m$, $n$, and $o$ of the $M_{\Lambda\bar{\Lambda}}$,
$M_{\bar{\Lambda} \eta}$ and $M_{\bar{\Lambda}\eta}$ distributions in
data and PHSP $\mathrm{MC}$ sample.  The weighted MC sample is iteratively
weighted twice in the same way to achieve better agreement between
data and MC simulation. As shown in Fig.~\ref{fig:mcdataComp}, the
distributions of the weighted PHSP MC sample~(hatched histograms) are
consistent with the background-subtracted data~(black points). The
efficiencies from the weighted PHSP MC samples are calculated as

\begin{equation}
  \label{eq:EffRewighted}
  \varepsilon^\text{wtd}=\frac{\sum^{N_{\mathrm{rec}}}_{i=1}{w^{i}(M_{\Lambda \bar{\Lambda}}, M_{\Lambda \eta}, M_{\bar{\Lambda}
    \eta})}}{\sum^{N_{\mathrm{gen}}}_{j=1}{w^{j}(M_{\Lambda \bar{\Lambda}}, M_{\Lambda \eta}, M_{\bar{\Lambda}
    \eta})}}, 
\end{equation} 
where $N_{\mathrm{rec}}$ and $N_{\mathrm{gen}}$ are the numbers of
reconstructed events and generated events and $w^{i~(j)}(M_{\Lambda
  \bar{\Lambda}}, M_{\Lambda \eta}, M_{\bar{\Lambda}
      \eta})$ is the weighting factor for event $i$~($j$).

In Eq.~(\ref{eq:MCweightingfactor}), data is used in the
calculation of the weighting factor. The statistical uncertainties of
the numbers of data events in the intervals of the
background-subtracted data distributions
propagate to the weighting factor and contribute to the uncertainty
of the efficiency~[see Eq.~(\ref{eq:EffRewighted})] which will be
considered in Section~\ref{sec:sysErr}.

To evaluate $\varepsilon$ after the ISR correction and
$(1+\delta)$, an iterative procedure is
performed~\cite{sun2020iterative}. The initial value of
$\varepsilon$ for the initial line shape is obtained by the weighted
PHSP MC sample. The initial
value of $(1+\delta)$ is determined by {\sc kkmc} with the line shape
assumed to be flat. The LOWESS method~\cite{Lowess_Cleveland1981} is
used to smooth the line shape. The iterative procedure is repeated
until the ratio of the measured cross section to the corresponding value
from previous iteration is consistent with one within the statistical
uncertainty. Table~\ref{tab:cswitherr} summarizes the $\varepsilon$ values
after ISR correction and $(1+\delta)$ for each energy point.

\begin{table*}[t]
  \centering
  \caption{Born cross sections at the thirty-one energy points for the process
    $e^+e^-\rightarrow\Lambda\bar{\Lambda}\eta$, where $\sqrt{s}
    $~\cite{BESIII:2020eyu, BESIII:2022dxl, BESIII:2022Lum} is the
    center-of-mass energy, $\mathcal{L}_\text{int}$ is the
    integrated luminosity~\cite{BESIII:2015qfd, BESIII:2022dxl, BESIII:2022Lum},
    signal yield is the number of signal events in the signal region where the uncertainty is
    statistical only, $\varepsilon$ is the reconstruction efficiency
    after the ISR correction where the uncertainty is from the
    limited data sample sizes, $(1+\delta)$ and $(1+\delta^{v})$ are
    the ISR correction factor and vacuum polarization factor,
    respectively, and $\sigma$ is the Born cross section where the
  first uncertainty is statistical and the second is
systematic.}\label{tab:cswitherr}
\begin{ruledtabular}
  \begin{tabular}{ccccccc}
    $\sqrt{s}$ (GeV) & $\mathcal{L}_\text{int}$ ($\mathrm{pb}^{-1}$)& Signal yield& $\varepsilon$ (\%)& ($1+\delta$) & ($1+\delta^{v}$) & $\sigma$ ($\mathrm{pb}$)\\
  \hline
 3.5106& ~458.7&~80.3$\pm$11.0&$15.6 \pm0.1$& 0.864&1.045 &$3.13\pm 0.43\pm 0.17$\\
 3.7730& 2916.9&378.9$\pm$24.1&$18.4 \pm0.2$& 0.948&1.059 &$1.77\pm 0.11\pm 0.10$\\
 3.8720& ~219.2&~15.8$\pm$ 5.1&$17.4 \pm0.8$& 0.991&1.050 &$1.00\pm 0.32\pm 0.07$\\
 4.0076& ~482.0&~64.1$\pm$ 9.0&$18.8 \pm0.5$& 0.979&1.046 &$1.77\pm 0.25\pm 0.10$\\
 4.1285& ~401.5&~36.4$\pm$ 7.0&$18.0 \pm0.5$& 1.054&1.053 &$1.17\pm 0.23\pm 0.08$\\
 4.1574& ~408.7&~37.1$\pm$ 7.0&$19.6 \pm0.6$& 1.057&1.054 &$1.09\pm 0.21\pm 0.07$\\
 4.1784& 3189.0&277.4$\pm$19.5&$20.2 \pm0.3$& 1.054&1.055 &$1.00\pm 0.07\pm 0.06$\\
 4.1888& ~570.0&~39.8$\pm$ 7.5&$19.0 \pm0.6$& 1.083&1.057 &$0.82\pm 0.15\pm 0.05$\\
 4.1989& ~526.0&~63.2$\pm$ 8.8&$20.0 \pm0.5$& 1.055&1.057 &$1.38\pm 0.19\pm 0.08$\\
 4.2092& ~517.1&~42.9$\pm$ 7.7&$19.1 \pm0.6$& 1.039&1.057 &$1.00\pm 0.18\pm 0.06$\\
 4.2187& ~569.2&~47.0$\pm$ 7.8&$17.7 \pm0.6$& 1.097&1.057 &$1.06\pm 0.17\pm 0.07$\\
 4.2263& 1100.9&~80.6$\pm$10.3&$20.7 \pm0.5$& 1.048&1.057 &$0.81\pm 0.10\pm 0.05$\\
 4.2357& ~530.3&~41.5$\pm$ 7.5&$20.3 \pm0.6$& 1.060&1.055 &$0.89\pm 0.16\pm 0.06$\\
 4.2438& ~538.1&~48.6$\pm$ 8.0&$20.0 \pm0.7$& 1.085&1.056 &$1.01\pm 0.17\pm 0.07$\\
 4.2580& ~828.4&~58.7$\pm$ 8.9&$19.9 \pm0.6$& 1.060&1.054 &$0.79\pm 0.12\pm 0.05$\\
 4.2668& ~531.1&~39.1$\pm$ 7.4&$19.1 \pm0.8$& 1.029&1.053 &$0.91\pm 0.17\pm 0.07$\\
 4.2879& ~502.4&~46.9$\pm$ 7.5&$20.9 \pm0.6$& 1.000&1.053 &$1.05\pm 0.17\pm 0.07$\\
 4.3121& ~501.2&~40.0$\pm$ 7.3&$18.6 \pm0.7$& 1.026&1.052 &$1.04\pm 0.19\pm 0.07$\\
 4.3374& ~505.0&~44.6$\pm$ 7.4&$20.0 \pm0.6$& 1.020&1.051 &$1.06\pm 0.17\pm 0.07$\\
 4.3583& ~543.9&~53.6$\pm$ 7.8&$20.3 \pm0.6$& 1.058&1.051 &$1.11\pm 0.16\pm 0.07$\\
 4.3774& ~522.7&~27.1$\pm$ 6.0&$20.4 \pm0.7$& 1.114&1.051 &$0.58\pm 0.13\pm 0.04$\\
 4.3964& ~507.8&~25.2$\pm$ 5.9&$17.7 \pm0.7$& 1.141&1.052 &$0.58\pm 0.13\pm 0.05$\\
 4.4156& 1090.7&~72.1$\pm$ 9.8&$20.5 \pm0.6$& 1.092&1.053 &$0.72\pm 0.10\pm 0.05$\\
 4.4400& ~569.9&~36.5$\pm$ 7.0&$19.7 \pm0.6$& 0.943&1.055 &$0.81\pm 0.15\pm 0.08$\\
 4.4671& ~111.1&~13.4$\pm$ 3.9&$23.0 \pm1.1$& 0.913&1.055 &$1.36\pm 0.39\pm 0.13$\\
 4.5995& ~586.9&~36.4$\pm$ 6.7&$21.3 \pm0.8$& 1.122&1.055 &$0.61\pm 0.11\pm 0.05$\\
 4.6280& ~521.5&~31.5$\pm$ 6.3&$18.5 \pm0.7$& 1.133&1.055 &$0.69\pm 0.14\pm 0.05$\\
 4.6612& ~529.6&~31.6$\pm$ 6.1&$20.3 \pm0.9$& 1.119&1.056 &$0.48\pm 0.12\pm 0.04$\\
 4.6409& ~552.4&~22.9$\pm$ 5.5&$17.1 \pm0.6$& 1.205&1.056 &$0.63\pm 0.12\pm 0.04$\\
 4.6819& 1669.3&~74.0$\pm$ 9.6&$17.2 \pm0.5$& 1.258&1.056 &$0.49\pm 0.06\pm 0.03$\\
 4.6988& ~536.5&~23.1$\pm$ 5.5&$19.2 \pm0.8$& 1.188&1.056 &$0.45\pm 0.11\pm 0.04$\\
 \end{tabular}
\end{ruledtabular}
\end{table*}                       

\begin{figure*}[t]
  \centering
  \includegraphics[width=1\textwidth]{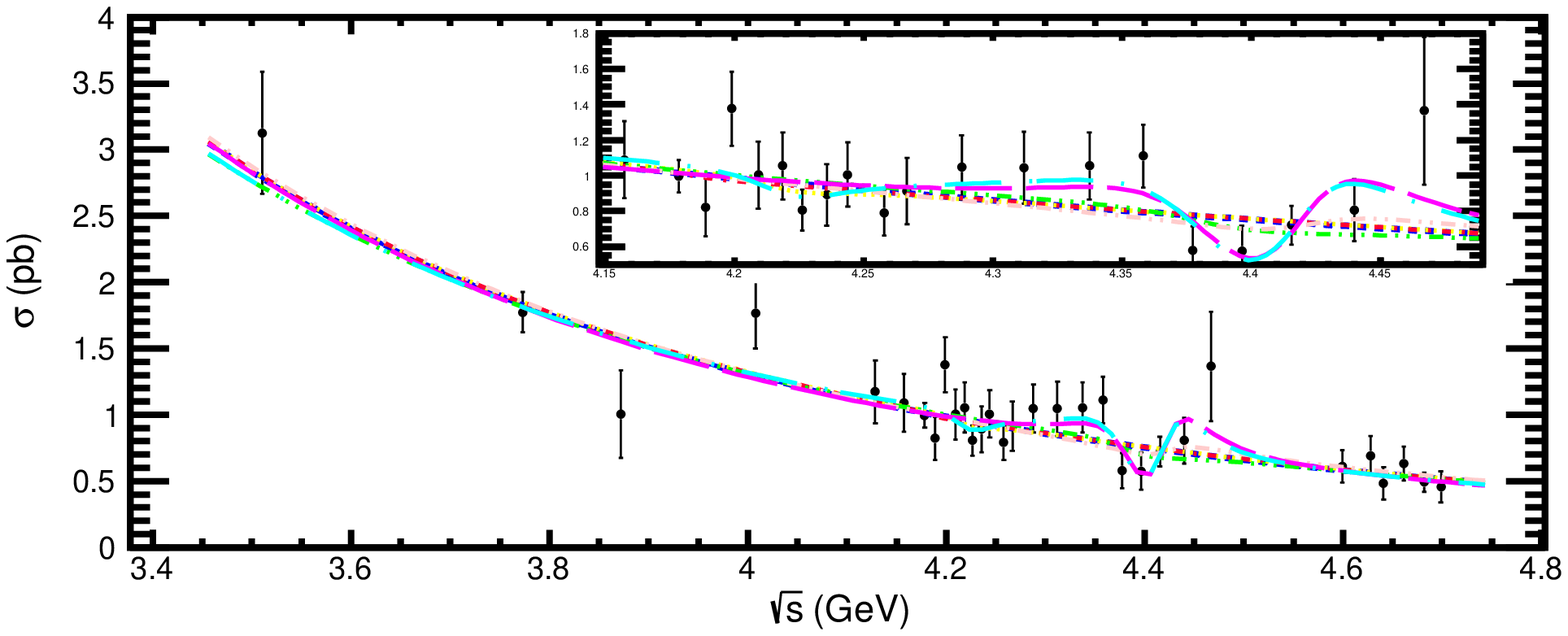}
  \caption{Measured Born cross sections of the process
    $e^+e^-\rightarrow\Lambda\bar{\Lambda}\eta$ at $\sqrt{s}$
    from 3.5106 to 4.6988 GeV indicated by the black points with
    error bars~(combined statistical and systematic
    uncertainties). The dashed blue line is the fit result for
    the fit model with the power-law function only. The solid
    red, yellow, green, and pink lines are the fits
     with a coherent sum of the continuum amplitude
    and Breit-Wigner amplitude of resonances $\psi(4160)$,
    $\psi(4220)$, $\psi(4360)$, and $\psi(4415)$, respectively. The
    solid magenta
    line denotes the fit with a coherent sum of the continuum
    amplitude, the resonance amplitude of resonance $\psi(4360)$,
    and the resonance amplitude resonance for
  $\psi(4415)$. The solid cyan line shows the fit with a coherent sum
  of a continuum amplitude and three resonance amplitudes for
  $\psi(4220)$, $\psi(4360)$, and $\psi(4415)$. The subplot shows the expanded region where the resonances are
expected.}\label{fig:csFitWithOneReso}
\end{figure*}

Table~\ref{tab:cswitherr} also summarizes the measured Born cross sections
at the thirty-one energy points. To search for possible resonances
decaying into $\Lambda\bar{\Lambda}\eta$, the least-squares method is
used to perform the fit of the Born cross sections with both
statistical and systematic uncertainties taken into account, ignoring
the correlation between the different energy points. The continuum production
of $e^+e^-\to\Lambda\bar{\Lambda}\eta$ is described by a power-law
function $\sigma_{\mathrm{con}}=C/s^{\lambda}$~\cite{ppPi0_201745}.
The possible resonances are added coherently as

\begin{equation}
  \label{eq:BWCon}
  \begin{aligned}
    \sigma (s)  = & \left|  \sqrt{\sigma_{\mathrm{con}}}\right. \\ 
                &+ \sum_k
  \left.\sqrt{\sigma_{\mathrm{Reso}}^k} e^{i \phi^k} \frac{\mathit{m_k} \Gamma_k}
  {s - m_k^2 + i \mathit{m_k} \Gamma_k}^{}\right|^2 ,
  \end{aligned}
\end{equation}
where $m_k$ and $\Gamma_k$ are mass and width of the resonance $k$,
$\sigma_{\mathrm{Reso}}^k$ is production cross section of the resonance $k$ at
the $m_k$, $\phi^k$ is the phase angle. Since most of the data samples are
taken at $\sqrt{s}$
from 4.2 to 4.6 GeV, as shown in Table~\ref{tab:cswitherr}, the four
well-established resonances, $\psi(4160)$, $\psi(4220)$, $\psi(4360)$,
and $\psi(4415)$ are taken into account in the fit. The masses and
widths of these resonances are fixed to the corresponding values in the
PDG~\cite{ParticleDataGroup:2020ssz} and other parameters are free.
Figure~\ref{fig:csFitWithOneReso} shows the fits to the Born cross section
with various fit strategies. 
The fit with the power-law function yields the goodness of fit
$\chi^2/\text{ndf}=30.3/29$, where ndf denotes the
number of degrees of freedom. The fitted parameters $C$ and $\lambda$
are $(4.1\pm 2.5)\times 10^3$ GeV$^{2\lambda}$~pb and
$2.9\pm 0.2$, respectively. To search for the potential resonance, we perform
fit using coherent sum of the continuum amplitude and one resonance amplitude.
The fits with hypotheses of $\psi(4160)$, $\psi(4220)$, $\psi(4360)$, and
$\psi(4415)$ give $\chi^2/\text{ndf}$ of 30.2/27, 29.9/27, 29.4/27, and
29.6/27, respectively. Because of the two floated
parameters~($\sigma_\text{Reso}^k$ and $\phi^k$), the ndf for each fit is
reduced by two compared to the fit with the power-law function only.
The significance for each hypothesis with the resonance is less than
1$\sigma$, which is calculated with the changes of ndf and $\chi^2$. According
to the goodness of fit, it is preferable to
describe the cross section line shape by the power-law function. The
power-law function, however, cannot describe the region around $4.4\mathrm{~GeV}$. Adding a single resonance amplitude cannot model the region
well, either. Therefore, we fit the line
shape with a coherent sum of a continuum amplitude, a resonance
amplitude of $\psi(4360)$, and a resonance amplitude of $\psi(4415)
$, shown as the magenta line in Fig.~\ref{fig:csFitWithOneReso}. The
fit with the two resonances delivers a better description around 4.4 GeV
with a goodness of fit equal to 24.12/25 with corresponding $p$ value of
0.512. The global significance
with respect to the fit with the power-law function is 1.3$\sigma$.
Since the cross section at 4.1989 GeV deviates from the fit result by
more than 1.0$\sigma$, we also fit the cross section line shape with
a coherent sum of a continuum amplitude and three resonance
amplitudes of $\psi(4220)$, $\psi(4360)$, and $\psi(4415)$ shown as
the cyan line. The fit quality is 22.63/23 corresponding to $p$ value of
0.483. The fit quality is worse than the fit with
two resonances because of the increasing number of parameters.

\section{\boldmath Study of \texorpdfstring{$\Lambda\bar{\Lambda}$}
{MLambdaLambdabar} Mass Threshold Enhancement}%
\label{sec:StudyofMLambdabarLambda}

\begin{figure}[t]
  \centering
  \begin{overpic}[width=0.45\textwidth, percent]{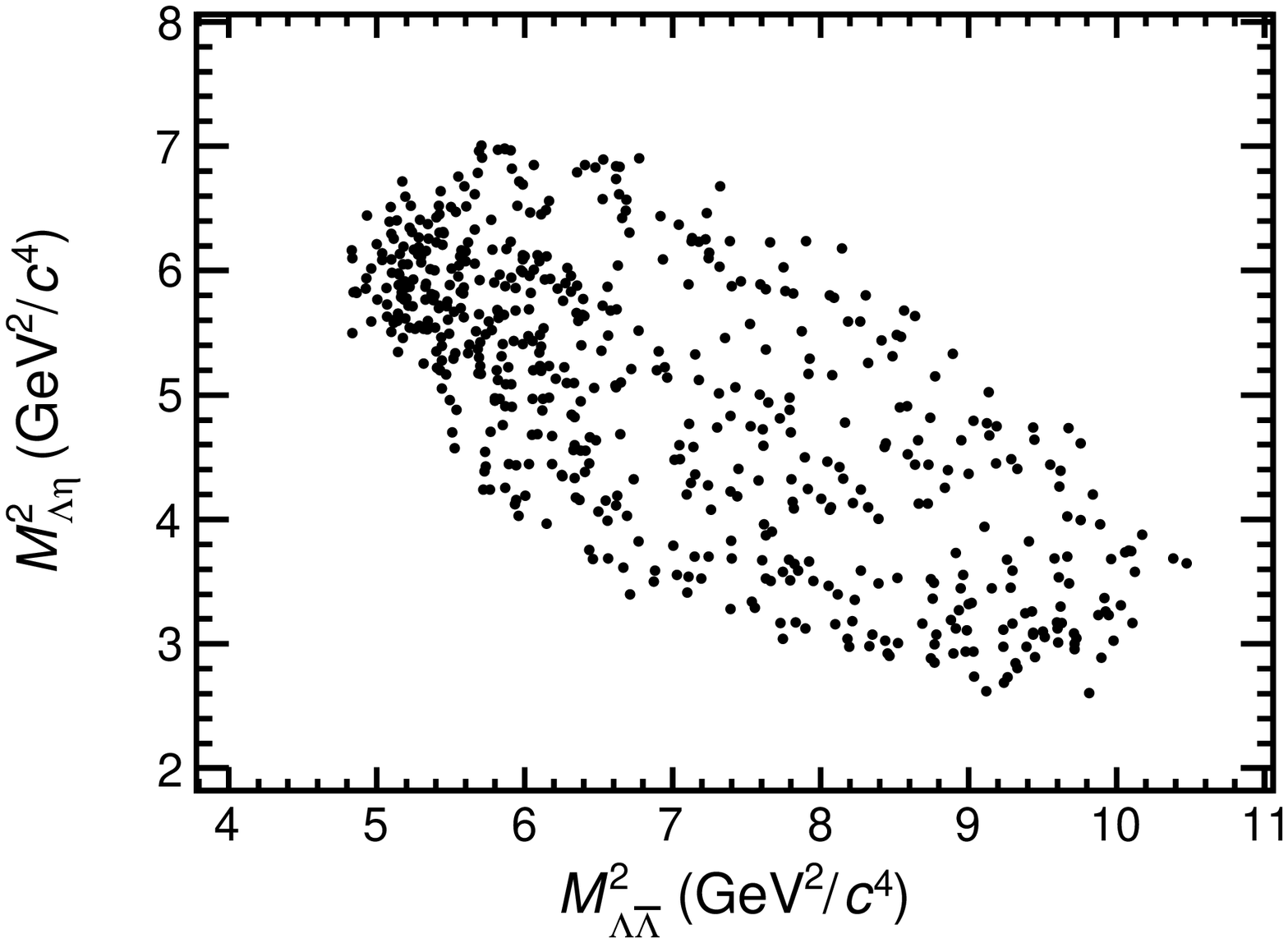}
    \put(77,60){(a)}
  \end{overpic}
  \begin{overpic}[width=0.45\textwidth, percent]{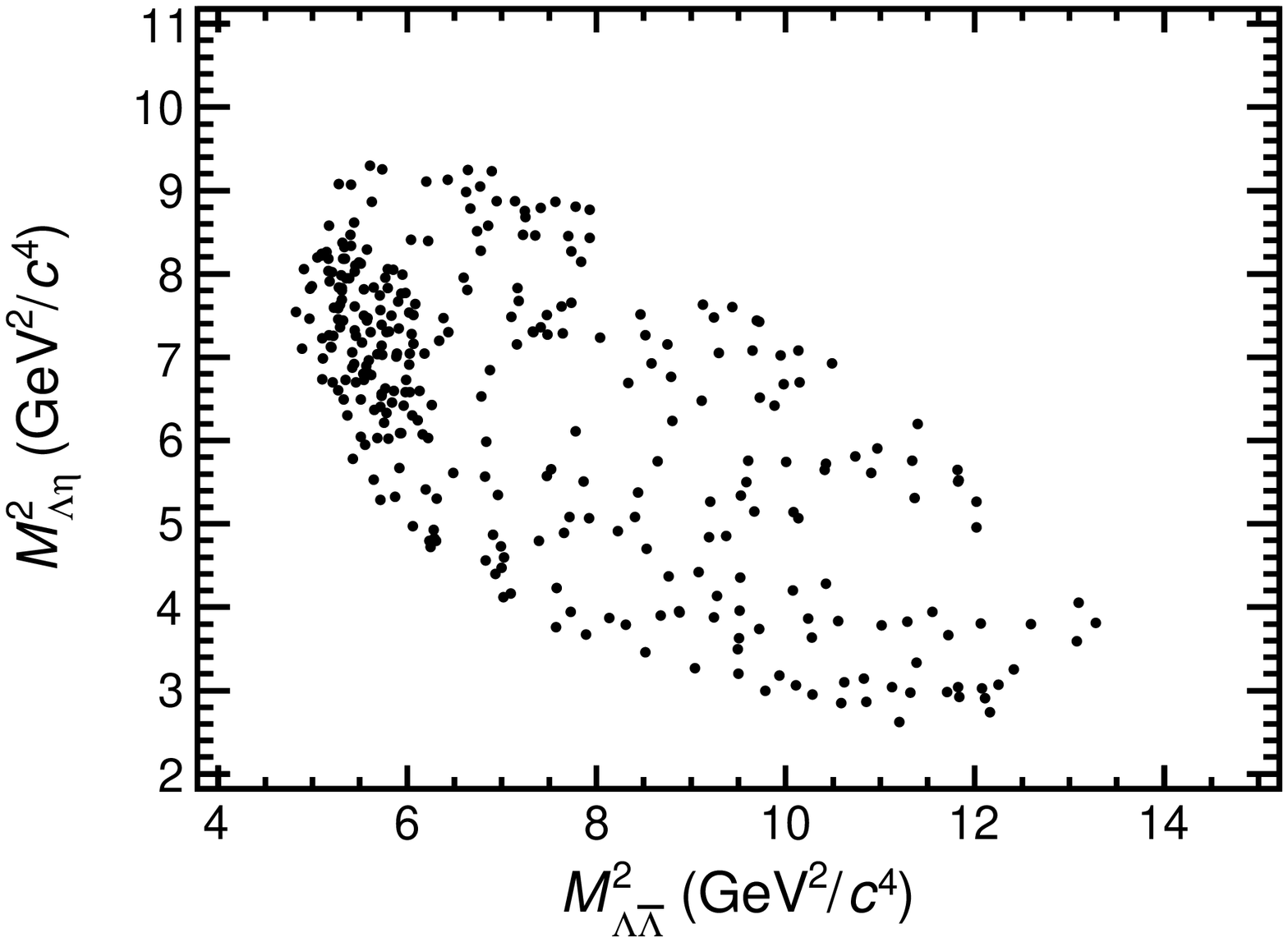}
    \put(77,60){(b)}
  \end{overpic}
  \caption{Dalitz plots of $M^2_{\Lambda\eta}$ versus $M^{2}
  _{\Lambda\bar{\Lambda}}$ in data taken at (a) $\sqrt{s}=3.7730$ GeV 
and (b) $\sqrt{s}=4.1784$ GeV.}\label{fig:dalitz}
\end{figure}

\begin{figure}[b]
  \centering
  \includegraphics[width=0.45\textwidth]{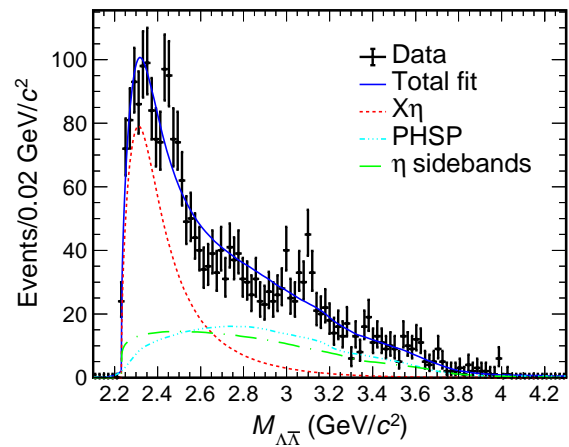}
  \caption{Sum of the simultaneous fits to the $M_{\Lambda\bar{\Lambda}}$
  distributions at the thirty-one energy points. The black points with error
bars are data. The solid blue curve shows the total fit. The dashed
red line is the enhancement. The cyan line shows PHSP component. The
dashed green line denotes the background
component.}\label{fig:simfit}
\end{figure}

A clear enhancement is observed near the $\Lambda\bar{\Lambda}$
mass threshold as shown in Fig.~\ref{fig:simfit}.
This structure cannot be described by the
PHSP MC simulation~(the hatched histograms in 
Fig.~\ref{fig:mcdataComp}). As shown in Fig.~\ref{fig:dalitz},
examinations of the Dalitz plots of the process
$e^+e^-\rightarrow\Lambda\bar{\Lambda}\eta$ indicate that the
threshold enhancement in ${\Lambda\bar{\Lambda}}$ mass spectra is the
dominant component. In addition, a small peak from the process $e^+e^-\rightarrow
J/\psi\eta\rightarrow\Lambda\bar{\Lambda}\eta$ at 3.1 GeV/$c^2$ can be seen in Fig.~\ref{fig:mcdataComp}.

To determine the mass and width of the enhancement, a simultaneous
unbinned maximum-likelihood fit is performed to the
$\Lambda\bar{\Lambda}$ invariant mass spectra of the data samples at
the thirty-one energy points. The enhancement is modeled by a
Breit-Wigner function with mass-dependent width of the following
form~\cite{BESIII:2021fqx}:

\begin{equation}
\begin{aligned}
\mathrm{d} N / \mathrm{d} M_{\Lambda \bar{\Lambda}} \propto &
\varepsilon(M_{\Lambda \bar{\Lambda}})\left(p_{X}\right)^{2 l_{X
\eta}+1} f_{l_{X \eta}}^{2}\boldsymbol{(}\left(R
p_{X}\right)^{2}\boldsymbol{)} \\
& \times\left|\text{BW}\left(M_{\Lambda \bar{\Lambda}}\right)\right|^{2} \\
& \times\left(p_{\Lambda}\right)^{2 l_{\Lambda \bar{\Lambda}}+1}
f_{l_{\Lambda \bar{\Lambda}}}^{2}\boldsymbol{(}\left(R p_{\Lambda}\right)^{2}\boldsymbol{)}, 
\end{aligned}
\end{equation}
where $X$ denotes the threshold enhancement, the mass-dependent reconstruction
efficiency $\varepsilon(M_{\Lambda \bar{\Lambda}})$ is
determined by the MC simulation, $p_{X}$ is the momentum of $X$ in the
$e^+e^-$ center-of-mass system,
$l_{X\eta}$ is the orbital angular momentum between $X$ and
$\eta$, and $f_l(z)$ is the Blatt-Weisskopf barrier
factor~\cite{BlattWeiss_PhysRevLett.96.102002} with $f_0(z)=1$ and
$f_1(z)=\frac{1}{\sqrt{1+z}}$, where $l$ is the orbital angular
momentum. The barrier radius factor $R$ is assumed to be $1.0~(\mathrm{GeV}/c)^{-1}$ since it
is poorly known for baryons. The Breit-Wigner amplitude is defined as

\begin{equation}
  \label{eq:BW}
  \text{BW} \propto \frac{1}{M^2_X -M^2_{\Lambda \bar{\Lambda}}  - iM_X
  \Gamma},
\end{equation}
where $M_X$ is the mass of $X$. The mass dependent width $\Gamma$
is written as

\begin{equation}
  \label{eq:masswidth}
  \Gamma = \Gamma_X \left( \frac{p_{\Lambda}}{p^X_{\Lambda}} \right)
  ^{2 l_{\Lambda \bar{\Lambda}} + 1} \frac{M_X}{M_{\Lambda
  \bar{\Lambda}}} \frac{f^2_{l_{\Lambda \bar{\Lambda}}}
\boldsymbol{(}(Rp_{\Lambda})^2\boldsymbol{)}}{f^2_{l_{\Lambda \bar{\Lambda}}}
\boldsymbol{(}(Rp^X_{\Lambda})^2\boldsymbol{)}},
\end{equation}
where $\Gamma_X$ is the width of $X$, $p_\Lambda$ is
the momentum of $\Lambda$ in the rest frame of $X$, $p_\Lambda^X$ is the
momentum when $M_{\Lambda\bar{\Lambda}}=M_X$, and
$l_{\Lambda\bar{\Lambda}}$ is the orbital angular momentum between
$\Lambda$ and $\bar{\Lambda}$. In the simultaneous fit, $M_X$ and
$\Gamma_X$ are the common parameters. Based on the conservation of
angular momentum, parity, and charge conjugation, the
possible quantum numbers of the enhancement are listed in
Table~\ref{tab:JPC}, which
summarizes the possible combinations of $l_{\Lambda\bar{\Lambda}}$
and $l_{X\eta}$. We assume the structure has $J^{PC}$ equal to
$1^{--}$ and $l_{\Lambda\bar{\Lambda}}$ equal to 0 since the
enhancement is near the $\Lambda\bar{\Lambda}$ mass threshold.
\begin{table}[b]
  \centering
  \caption{Possible quantum numbers of $X$ considering
  conservation of angular momenta, parity, and charge
conjugation, where $S_{\Lambda\bar{\Lambda}}$ is the total
spin of $\Lambda\bar{\Lambda}$ system}\label{tab:JPC}
\begin{ruledtabular}
  \begin{tabular}{ccc}
    $S_{\Lambda\bar{\Lambda}}$/$l_{\Lambda\bar{\Lambda}}$ of $\Lambda\bar{\Lambda}$ &$J^{PC}$ of $X$ &  $l_{X\eta}$\\ 
    \hline
   0/1 &$1^{+-}$  & 0\\
   0/1 &$1^{+-}$  & 2\\
   1/0 &$1^{--}$  & 1\\
   1/2 &$1^{--}$  & 1\\
   1/2 &$2^{--}$  & 1\\
  \end{tabular}
\end{ruledtabular}
\end{table}

\begin{figure}[t]
  \centering
  \includegraphics[width=0.45\textwidth]{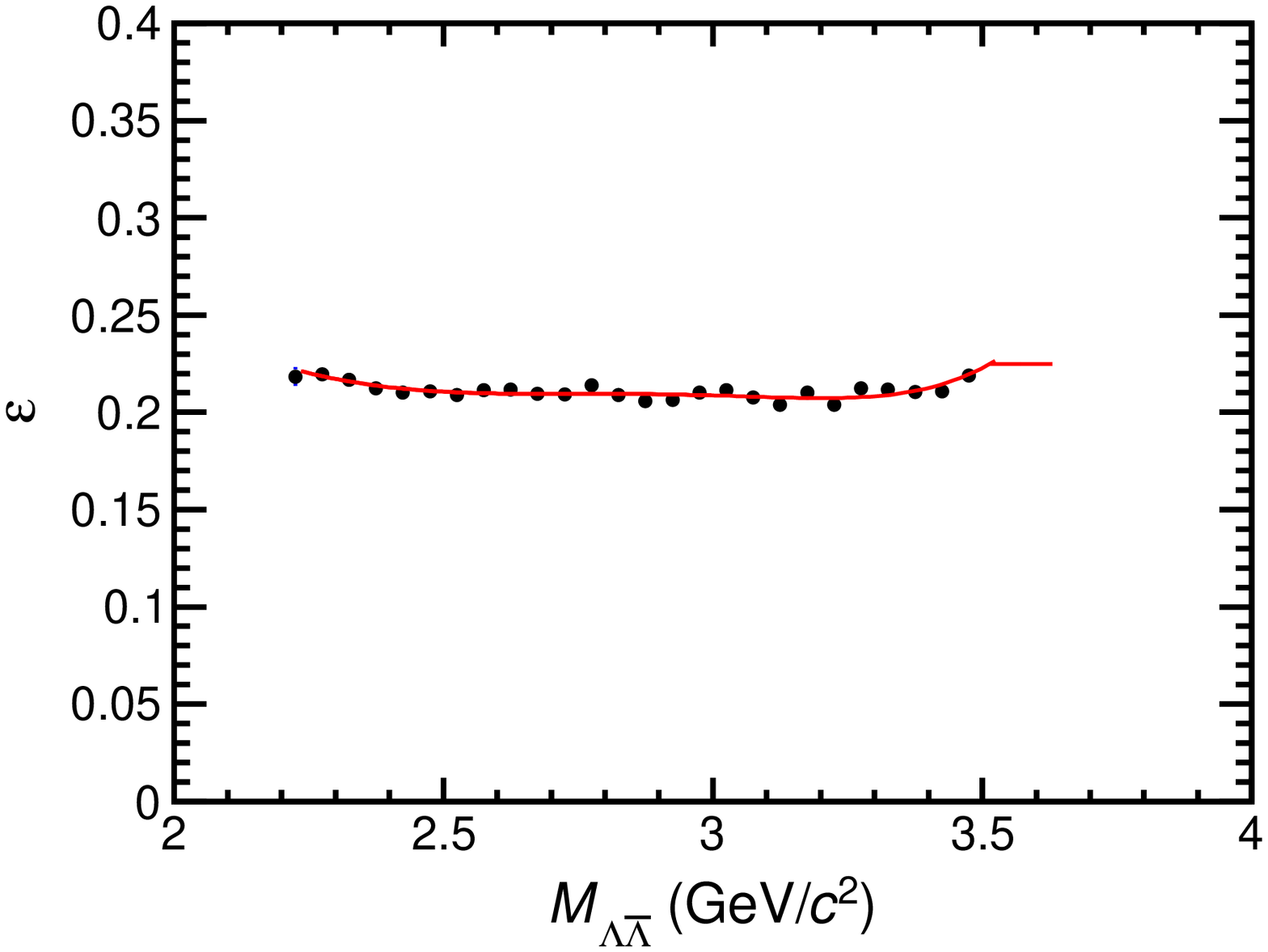}
  \caption{Fit of $\varepsilon(M_{\Lambda\bar{\Lambda}}) $ to the mass-dependent efficiency~(the black points) at
    $\sqrt{s}=4.1784$ GeV. The red line in the region
        $M_{\Lambda\bar\Lambda}<$ 3.5 GeV/$c^2$ represents  
  the fit result which is a six-order polynomial function. The red
  line in the region $M_{\Lambda\bar\Lambda}>$ 3.5 GeV/$c^2$ is the
extrapolation. The efficiency is obtained from the distribution of
the reconstructed value of $M_{\Lambda\bar{\Lambda}}$ divided by the
distribution of the generated value of
$M_{\Lambda\bar{\Lambda}}$.}\label{fig:MLLeff}
\end{figure}

We use MC simulations of the process $e^+e^-\rightarrow
\Lambda\bar{\Lambda}\eta$ to determine $\varepsilon(M_{\Lambda\bar{\Lambda}})$, where
the mass distribution of $\Lambda\bar{\Lambda}$ system follows a Breit-Wigner function with a relatively large width so that the
$M_{\Lambda\bar{\Lambda}}$ distribution is uniform within the
allowed region. The quantum numbers are configured according to the assumption
in the previous paragraph. Figure~\ref{fig:MLLeff} shows
$\varepsilon(M_{\Lambda\bar{\Lambda}})$ as a function of
$M_{\Lambda\bar{\Lambda}}$ at $\sqrt{s}=4.1784 \text{~GeV}$ with
$M_{\Lambda\bar{\Lambda}}<3.5$ GeV/$c^2$. We parametrize
$\varepsilon(M_{\Lambda\bar{\Lambda}})$ by a six-order polynomial
function. Because of the detector resolution, the reconstructed
distributions exceed the allowed region. As a result, the efficiency
cannot be determined correctly in the range of
$M_{\Lambda\bar{\Lambda}}>3.5$ GeV/$c^2$, which is not included in
the parametrization. Instead, the efficiency in this region, which is
far away from the threshold enhancement, is
estimated by a flat extrapolation assuming the efficiency is constant
in the region.

The events from the nonresonant process
$e^+e^-\rightarrow\Lambda\bar{\Lambda}\eta$ are described by shapes
derived from the PHSP MC samples. The background events are modeled by
the formula~\cite{Ablikim:2015swa}

\begin{equation}\label{eq:newbkgshape}
  f(M_{\Lambda \bar{\Lambda}}) \propto \left(M_{\Lambda \bar{\Lambda}}-M_{\mathrm{\min}}\right)^l\left(M_{\mathrm{\max}}
    -M_{\Lambda \bar{\Lambda}}\right)^h, 
\end{equation}
where 
$M_{\mathrm{\min}}=2M_{\Lambda}$ is the minimum allowed
$M_{\Lambda\bar{\Lambda}}$, $M_{\mathrm{\max}}=\sqrt{s}-M_\eta$ is
the maximum allowed $M_{\Lambda\bar{\Lambda}}$ calculated with
center-of-mass energy $\sqrt{s}$ and known mass of $\eta$,
$M_\eta$~\cite{ParticleDataGroup:2020ssz}, and $l$ and $h$ are
determined by fitting the $M_{\Lambda\bar{\Lambda}}$ spectra of
events from the $\eta$ sidebands. 
The magnitudes of the background shapes are fixed to 
the number of the fitted background events within the $\eta$ signal
region.

Figure~\ref{fig:simfit} shows the sum of the simultaneous fit to the
$M_{\Lambda\bar{\Lambda}}$ spectra at the thirty-one energy points.
The mass and width of $X$ obtained from the fit are
$(2356\pm7)\text{~MeV}/c^2$ and $(304\pm28)$ MeV, respectively,
where the uncertainties are statistical only. In the fit, the excited
$\Lambda$ states are not included because they are not significant in the
data~(\cref{fig:dalitz}) and are not
well-established~\cite{ParticleDataGroup:2020ssz}. Recently, an improved
measurement of $\Lambda(1670)$ via the $\psi(2S) \rightarrow \Lambda
\bar{\Lambda} \eta$ decay has been reported by the BESIII
collaboration~\cite{BESIII:2022cxi}. The effect of the $\Lambda(1670)$ on the
determination of mass and width will be discussed in \cref{subsec:sysSimFit}
for the estimation of systematic uncertainties. Because of the limited sample
sizes, the potential interferences among the threshold enhancement, excited
$\Lambda$ states, and three-body PHSP component are neglected. The combined
statistical signal significance is greater than $10\sigma$, which is
determined by the difference of likelihood values between the fit
with and without the model for the
enhancement~\cite{zhu2008statistical}.

\begin{figure}[b]
  \centering
  \includegraphics[width=0.45\textwidth]{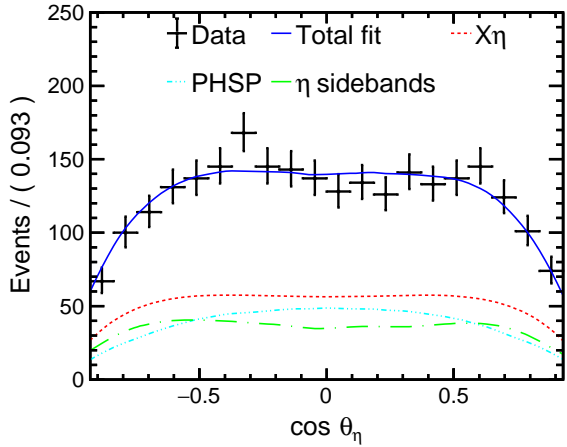}
  \caption{Sum of the simultaneous fits to the $\cos\theta$
    distributions of $\eta$ candidates at the thirty-one energy
    points.  The black points with error bars are the data. The solid
    blue line shows the total fit. The dashed red line shows the $X$
    component, and the dashed cyan line denotes the PHSP
    component. The dashed green line is the
    background.}\label{fig:angfitSim}
\end{figure}

In the determination of mass and width of $X$, $l_{X\eta}$
is assumed to be one. The $\cos\theta$ distributions of $\eta$
candidates should follow the distribution of
$\left(1+\cos^2\theta\right)$ when $l_{X\eta}=1$. Therefore, the angular
distributions of $X$ decays are studied to verify this assumption. We
perform a simultaneous unbinned
maximum-likelihood fit to the $\cos\theta$ distributions at the
thirty-one center-of-mass energies. The $\cos\theta$ distributions of
$X$ decays are modeled by the function $\varepsilon(\cos\theta)
\left(1+\alpha\cos^2\theta\right)$, where the parameter $\alpha$ is
free and taken as a common parameter. The reconstruction efficiency
$\varepsilon(\cos\theta)$ is parametrized by a fourth-order polynomial
function. The shape of the PHSP process is modeled by the MC simulations.
The background contributions are described by the shapes derived from
the $\eta$ sidebands. In the fit, the numbers of events for
the three components are fixed to the values obtained from the
simultaneous fit to the $M_{\Lambda\bar{\Lambda}}$ distributions.
Figure~\ref{fig:angfitSim} shows the sum of the simultaneous fit
to the distributions of $\cos\theta$. The parameter $\alpha$ is
determined to be $0.8 \pm0.3$, which is consistent with our
assumption of $\alpha=1$ considering the statistical uncertainty.

\section{Systematic Uncertainties}\label{sec:sysErr}
\subsection{Born cross section measurement}\label{subsec:sysErrCS}

According to Eq.~(\ref{eq:csborn}), the uncertainties on the cross
section measurement are associated with the luminosity, branching
fractions, reconstruction efficiency, fit procedure, ISR correction,
and vacuum polarization factor. The uncertainty of the reconstruction
efficiency includes contributions from the $\Lambda$ reconstruction,
PID of the pion, photon reconstruction, shower requirements, $\Sigma^0$
mass window, and weighting procedure~(limited data samples sizes, binning of data
distributions, $\eta$ mass window, and $\eta$ sidebands). The
uncertainty of the fit procedure incorporates the fit range,
background shape, peaking background, and signal shape. To reduce the
effect of statistical uncertainty due to limited sample sizes, we take
the samples at 3.7730 and 4.1784 GeV~(the two samples with the
highest luminosities) to estimate the uncertainties from the $\Lambda$
reconstruction, $\Sigma^0$ mass window, weighting of PHSP MC
samples~(except for the limited data sample size), and fit procedure.

\begin{enumerate}
  \item \textit{Luminosity}. The uncertainties from the integrated
    luminosities measured using the large-angle Bhabha events are assigned to
    be 1.0\%~\cite{BESIII:2015qfd, BESIII:2022dxl, BESIII:2022Lum}. 

\item \textit{Branching fractions}.
The systematic uncertainty due to the branching fractions 
$\mathcal{B}(\eta\rightarrow\gamma\gamma)$ and $\mathcal{B}(\Lambda\rightarrow p\pi^-)$ are
0.5\% and 0.8\%, respectively, taken from the
PDG~\cite{ParticleDataGroup:2020ssz}.

\item \textit{$\Lambda$ reconstruction}.
The systematic uncertainty from the $\Lambda$~($\bar{\Lambda}$)
reconstruction efficiency is estimated with the
equation~\cite{LamSysErrWeightAblikim:2018byv}

\begin{equation}
  \label{eq:SysErrLamReocn} \left| \frac{\Sigma_i 
  \frac{N_i}{\varepsilon_i^{\mathrm{Data}}} - \Sigma_i 
  \frac{N_i}{\varepsilon_i^{\mathrm{MC}}}}{\Sigma_i 
  \frac{N_i}{\varepsilon_i^{\mathrm{MC}}}} \right|,
\end{equation}
where $N_i$ is the number of data events in the interval $i$ with
background subtracted using the $\eta$ sidebands, and
$\varepsilon_i^\mathrm{Data}$ and $\varepsilon_i^\mathrm{MC}$ are the
efficiencies of $\Lambda$~$(\bar{\Lambda})$ reconstruction for the
data and MC simulations in interval $i$ of the two-dimensional~(momentum
and $\cos\theta$) distributions. The efficiencies are determined by a control sample of
$J/\psi\rightarrow\overline{p} K^{+} \Lambda+\text{c}.\text{c}
$.~\cite{LamErr_PhysRevLett.121.062003}. It takes the tracking of
proton and pion, PID of proton, mass window for
$\Lambda$~$(\bar{\Lambda})$ candidates, and the requirement on the
decay length into account. The uncertainty is estimated to be 2.8\%.

\item \textit{PID of pion}.
The uncertainty introduced by the PID of the pion is estimated to
be 1.0\% by a control sample of ${e}^{+} {e}^{-} \rightarrow {K}^{+} {K}
^{-} \pi^{+} \pi^{-}$~\cite{ablikim2021observation}.

\item \textit{Photon reconstruction}.
A systematic uncertainty of 1.0\%~\cite{PhotonEff} is assigned to the 
photon reconstruction efficiency. Hence, 2.0\% is taken as the
systematic uncertainty for two photons.

\item \textit{Shower requirements}.
The uncertainty associated with the requirements on the lateral
moment and $E_{3\times3}/E_{5\times5}$ is
0.8\%~\cite{sysErrShowerShape_Ablikim:2017ors}.

\item \textit{$\Sigma^0$ mass window}. The requirement
  $M_{\Lambda\gamma_{L}~(\bar{\Lambda}\gamma_{L})} \notin [1.113,
  1.273]$ GeV/$c^2$ is applied to suppress the background processes with
  $\Sigma^0$~$(\bar{\Sigma}^0)$ in the final state. By changing the width of the
  nominal mass window by $\pm$5 MeV/$c^2$, the corresponding
  uncertainty is estimated to be 0.8\%.

\item \textit{Weighting procedure}. Using the
  $M_{\Lambda\Bar{\Lambda}}$, $M_{\Lambda\eta}$, and $M_{\Bar{\Lambda}
  \eta}$ distributions from the data events, the PHSP MC is weighted
  to determine the reconstruction efficiency.
 
  \begin{enumerate}
\item \textit{Limited data sample size}.
 To estimate the uncertainties caused by the statistical fluctuation
 of the background-subtracted data distributions, we first produce a
 series of $M_{\Lambda\bar{\Lambda}}$, $M_{\Lambda\eta}$, and
 $M_{\bar{\Lambda}\eta}$ distributions for each energy point by
 sampling over the three background-subtracted data distributions with
 Poisson functions.  The means of the Poisson functions are the number
 of events in the intervals of the three background-subtracted data
 distributions. The PHSP MC samples are then weighted by the generated
 $M_{\Lambda\bar{\Lambda}}$, $M_{\Lambda\eta}$, and
 $M_{\bar{\Lambda}\eta}$ distributions. The resulting efficiency
 distribution for each energy point is fitted with a double-shouldered
 crystal ball function~\cite{das2016simple}. The fitted means and
 widths are taken as efficiencies and corresponding uncertainties from
 limited data sample sizes, respectively.

 \item \textit{Binning of data distributions}.
The widths of intervals in the $M_{\Lambda\Bar{\Lambda}}$,
$M_{\Lambda\eta}$, and $M_{\Bar{\Lambda}\eta}$
distributions are increased by 10\% to estimated corresponding
uncertainties. The resulting relative difference in the
efficiency is assigned as corresponding uncertainty, which is 0.5\%.

\item \textit{$\eta$ mass window}.
The width of the $\eta$ mass window is varied by $\pm$16
MeV/$c^2$~(twice the
resolution of the invariant mass of the $\eta$ candidates) to
estimate the corresponding uncertainty, which is 1.0\%.

\item \textit{$\eta$ sidebands}. To estimate the uncertainty due to
  the sidebands, their distances to the signal region  are varied by
  $\pm8$ MeV/$c^2$. The corresponding uncertainty is found to be negligible.

\end{enumerate}

\item \textit{Fit procedure}. The signal yield is determined by an
  unbinned maximum-likelihood fit to the $M_{\gamma\gamma}$ spectrum. The
  following aspects are considered when evaluating the systematic
  uncertainties associated with the fit procedure.
  \begin{enumerate}
    \item \textit{Fit range}.
      The ranges of fits to the $M_{\gamma\gamma}$ distributions are
      varied by $\pm100$ MeV/$c^2$. The resulting relative difference
      in the signal yield is taken as the corresponding uncertainty.
      The uncertainty is taken as 0.9\%.
\item \textit{Background shape}.
The nominal background shape, a linear function, is replaced by a
second-order Chebychev polynomial. The estimated uncertainty is
1.6\%.

\item \textit{Peak background}.
  For the peaking background from the processes
  $e^+e^-\rightarrow\eta\Sigma^{0}\bar{\Lambda}+\text{c}.\text{c}$.
  $\mathrm{and}$ $e^+e^-\rightarrow\eta\Sigma^0\bar{\Sigma} ^0$, we add
  the shapes of the two processes obtained from the MC simulations to
  the nominal model. The corresponding uncertainty is estimated to be
  2.9\%.

\item \textit{Signal shape}.
  For the signal shape, we set the fixed width of the Gaussian
  function to be free in the nominal fit. The corresponding
  uncertainty is estimated to be 0.5\%.

  \end{enumerate}
\item \textit{ISR correction}. For uncertainties from ISR
  correction factors, we replace the smoothing method
  LOWESS~\cite{Lowess_Cleveland1981} by the power-law function $C/s^{\lambda}$,
  where the parameters $C$ and $\lambda$ are determined by fitting
  the measured Born cross section. The relative changes with respect
  to the nominal cross sections are assigned as the systematic
  uncertainties summarized in Table~\ref{tab:sysErrISR}.
\begin{table}[t]
  \centering
  \caption{Relative uncertainties~(\%) from ISR correction in the
    measurement of the Born cross section.}
  \label{tab:sysErrISR}
\begin{ruledtabular}
  \begin{tabular}{{cccccr}}
  $\sqrt{s}$ (GeV) & Uncertainty& & $\sqrt{s}$ (GeV) & Uncertainty\\ 
  \hline
  3.5106& 0.1 & &4.2879& 1.4 \\
  3.7730& 0.3 & &4.3121& 1.7 \\
  3.8720& 1.4 & &4.3374& 0.9 \\
  4.0076& 0.5 & &4.3583& 1.7 \\
  4.1285& 2.1 & &4.3774& 3.4 \\
  4.1574& 2.3 & &4.3964& 4.5 \\
  4.1784& 3.0 & &4.4156& 2.4 \\
  4.1888& 2.4 & &4.4400& 8.2 \\
  4.1989& 1.4 & &4.4671& 6.2 \\
  4.2092& 0.8 & &4.5995& 5.6 \\
  4.2187& 0.7 & &4.6280& 2.6 \\
  4.2263& 1.2 & &4.6612& 2.0 \\
  4.2357& 1.0 & &4.6409& 2.8 \\
  4.2438& 1.2 & &4.6819& 3.4 \\
  4.2580& 1.2 & &4.6988& 3.9 \\
  4.2668& 3.8 \\
  \end{tabular}
\end{ruledtabular}
\end{table}

\item \textit{Vacuum polarization factor}.
  The uncertainty introduced by the vacuum polarization factor is
  less than 0.1\%~\cite{WorkingGrouponRadiativeCorrections:2010bjp},
  which is negligible compared to other sources of uncertainties.
\end{enumerate}

Table~\ref{tab:sumsyserr} summarizes the uncertainties for all the samples.

\begin{table}[b]
  \centering
  \caption{Systematic uncertainties in the
    measurement of the Born cross section.}\label{tab:sumsyserr}
\begin{ruledtabular}
 \begin{tabular}{lc}
  Source & Uncertainty \\
  \hline
  Luminosity                                                                              & 1.0\%                                   \\
  $\mathcal{B}(\eta\rightarrow\gamma\gamma)$ and $\mathcal{B}(\Lambda\rightarrow p\pi^-)$ & 0.9\%                                \\
  $\Lambda$ reconstruction                                                                & 2.8\%                                \\
  PID of pion                                                                             & 1.0\%                                \\
  Photon reconstruction                                                                   & 2.0\%                                \\
  Shower requirements                                                                     & 0.8\%                                \\
  $\Sigma^0$ mass window                                                                  & 0.8\%                                \\
  Limited data sample size                                                                & see Table~\ref{tab:cswitherr}    \\
  Binning of data distributions                                                           & 0.5\%                                \\
  $\eta$ mass window                                                                      & 1.0\%                                \\
  $\eta$ sidebands                                                                        & negligible                          \\
  Fit range                                                                               & 0.9\%                                \\
  Background shape                                                                        & 1.6\%                                \\
  Peaking background                                                                      & 2.9\%                                \\
  Signal shape                                                                            & 0.5\%                                \\
  ISR correction                                                                          & see Table~\ref{tab:sysErrISR} \\
  \end{tabular}
\end{ruledtabular}
\end{table}

\subsection{Mass and width of the threshold enhancement}\label{subsec:sysSimFit}

The mass and width of the threshold enhancement in the
$\Lambda\bar{\Lambda}$ mass spectra are determined by the
simultaneous unbinned maximum-likelihood fit. The uncertainties
associated with the mass and width are assigned as follows:

\begin{enumerate}
  
  \item \textit{Description of the threshold enhancement}.
In the simultaneous fit to the $M_{\Lambda\bar{\Lambda}}$ distributions,
the barrier radius is assumed to be 1.0. To determine the uncertainties
from the assumption, the simultaneous fit is repeated with the
barrier radius being free, which results in a mass difference of 1
MeV/$c^2$ and a width difference of 4 MeV. The uncertainties due to
the parametrization of efficiency curves
$\varepsilon(m_{\Lambda\bar{\Lambda}})$ are estimated by
replacing the sixth-order polynomial function with a seventh-order
polynomial function. The resulting differences in the mass and width
are negligible.

\item \textit{Background shape.}
The uncertainties related to the background description are estimated
by changing Eq.~(\ref{eq:newbkgshape}) to the shapes of histograms
obtained from events in $\eta$ sidebands of the data. A fit under the scenario
yields a mass difference of 5 MeV/$c^2$ and a width difference of
11 MeV. 
\item \textit{$e^+e^- \to J/\psi\eta \to \Lambda\bar{\Lambda}\eta$.}
  In the $\Lambda\bar{\Lambda}$ mass spectrum shown in
  Fig.~\ref{fig:mcdataComp}, there is a small peak at the $J/\psi$ mass
  from the process $e^+e^- \to J/\psi\eta \to
  \Lambda\bar{\Lambda}\eta$. To estimate the effect of this process
  on the determination of mass and width of the threshold
  enhancement, we add the shape obtained from the MC simulation of
  $e^+e^- \to J/\psi\eta \to \Lambda\bar{\Lambda}\eta$ to the nominal
  fit model. The differences in the mass and width are $1
  \mathrm{~MeV}/c^2$ and $1\text{~MeV}$, respectively.

\item \textit{$e^+e^- \to \Lambda(1670)\bar\Lambda+\mathrm{c.c.} \to
  \Lambda\bar{\Lambda}\eta$.} To consider effect of the
  excited $\Lambda$
  state $\Lambda(1670)$ on the measurement of mass and width, we add the shape
  of the process obtained from the MC simulation to the nominal
  model. The resulting differences in mass
  and width, which are $11 \mathrm{~MeV}/c^2$ and 48 MeV, to the nominal results are
  assigned as corresponding uncertainties.

\item \textit{$\Sigma^0$ mass window, $\eta$ mass window, and $\eta$
  sidebands}.
  The $\Sigma^0$ mass window, $\eta$ mass window,
  and $\eta$ sidebands are altered as mentioned in
  Sec.~\ref{subsec:sysErrCS}, to estimate corresponding uncertainties. The
  maximum differences in mass are 6, 9, and 0.1
  MeV/$c^2$, and the maximum differences in width are 3, 20, and
  1 MeV, which are assigned as corresponding systematic
  uncertainties.

\end{enumerate}
After adding these uncertainties in quadrature, the total systematic
uncertainties on the mass and width of the enhancement are determined
to be 17 MeV/$c^2$ and $54 \text{~MeV}$, respectively.

\section{Summary}
Based on the thirty-one data samples taken at the center-of-mass
energies from 3.51 GeV to 4.70 GeV, we measure the Born cross
section of the process $e^+e^-\rightarrow\Lambda\bar{\Lambda}\eta$,
as shown in Table~\ref{tab:cswitherr}. No significant structure is
observed in the line shape of the Born cross section, which can
be described by a power-law function $C/s^{\lambda}$ with
$C=(4.1\pm 2.5)\times 10^3$ GeV$^{2\lambda}$pb and
$\lambda=2.9\pm 0.2$.

Further, a clear enhancement is observed near the
$\Lambda\bar{\Lambda}$ mass threshold. The contribution of the excited
$\Lambda$ are neglected in the study of
$M_{\Lambda\bar{\Lambda}}$ distribution since they are not significant in the
data and are not well-established~\cite{ParticleDataGroup:2020ssz}. Because of
the limited data sample sizes, the 
potential interferences among the structure, excited $\Lambda$ states, and
phase space process are also not considered. A simultaneous fit to the
$\Lambda\bar{\Lambda}$ mass spectra, assuming $J^{PC}=1^{--}$, yields
a mass of $(2356\pm7\pm17) \mathrm{~MeV}/c^2$ and width of
$(304\pm28\pm54)\text{~MeV}$ for the structure. The first
uncertainties are statistical and the second are systematic. The
statistical signal significance of the structure is larger than
10$\sigma$ over the hypothesis of the pure contribution from the PHSP
process of $e^+e^-\rightarrow\Lambda\bar{\Lambda}\eta$. The
$\cos\theta$ distributions of the structure can be described by a
function of $(1+\alpha \cos^2\theta)$. A simultaneous fit to the
$\cos\theta$ distributions gives $\alpha=0.8 \pm0.3$ which is
consistent with our assumption of $\alpha=1$, where the uncertainty is
statistical only. In the PDG, there is no well-established resonance
matching the structure in terms of the resonance parameters. A vector
hexaquark state decaying into $\Lambda\bar{\Lambda}$ is proposed with
a mass of 2200 MeV/${c}^2$ and a width of
$32\text{~MeV}$~\cite{Gerasyuta:2020fii}. The mass is close to our
measurement but the width of the hexaquark state is much narrower than
the width of the observed structure. In the framework of QCD sum rules, an
investigation of light baryonium states indicates that there exit possible
light baryonium states, including the $\Lambda\bar\Lambda$ state with $J^{PC}$
of $1^{--}$~\cite{Wan:2021vny}. The mass for the $\Lambda\bar\Lambda$ state is evaluated to be
$2.34\pm0.12$ GeV/$c^2$, which is consistent with our measurement. For the structure
with $M=2290
\pm 20 \mathrm{~MeV}/c^{2}$ and $\Gamma=275 \pm 35 \mathrm{~MeV}$ observed with a
partial wave analysis of PS185 data~\cite{Bugg:2004rj}, the width is
consistent with our result within the uncertainties, while the mass difference
cannot be covered by the uncertainties. For the recently observed
near-threshold enhancement in the process
$e^+e^-\rightarrow\phi\Lambda\bar{\Lambda}$~\cite{BESIII:2021fqx}, its
$C$ parity is opposite to the enhancement in our analysis. The
measured values of mass and width are $(2262\pm4\pm 28 )\text{
  MeV}/c^2$ and $( 72\pm 5 \pm43 )\text{~MeV}$, respectively, which
are smaller than the mass and width of our measurements. A future larger data sample~\cite{BESIII:2020nme} 
and theories incorporating a partial wave analysis could lead to
a better understanding of the observed structure.

\acknowledgments
The BESIII collaboration thanks the staff of BEPCII and the IHEP
computing center for their strong support. This work is supported in
part by National Key R\&D Program of China under Contracts Nos.
2020YFA0406300, 2020YFA0406400; National Natural Science Foundation
of China (NSFC) under Contracts Nos. 11635010, 11735014, 11835012,
11935015, 11935016, 11935018, 11961141012, 12022510, 12025502,
12035009, 12035013, 12192260, 12192261, 12192262, 12192263, 12192264,
12192265; the Chinese Academy of Sciences (CAS) Large-Scale
Scientific Facility Program; the CAS Center for Excellence in
Particle Physics (CCEPP); Joint Large-Scale Scientific Facility Funds
of the NSFC and CAS under Contract No. U1832207; CAS Key Research
Program of Frontier Sciences under Contracts Nos. QYZDJ-SSW-SLH003,
QYZDJ-SSW-SLH040; 100 Talents Program of CAS; The Institute of
Nuclear and Particle Physics (INPAC) and Shanghai Key Laboratory for
Particle Physics and Cosmology; ERC under Contract No. 758462;
European Union's Horizon 2020 research and innovation programme under
Marie Sklodowska-Curie grant agreement under Contract No. 894790;
German Research Foundation DFG under Contracts Nos. 443159800,
455635585, Collaborative Research Center CRC 1044, FOR5327, GRK 2149;
Istituto Nazionale di Fisica Nucleare, Italy; Ministry of Development
of Turkey under Contract No. DPT2006K-120470; National Science and
Technology fund; National Science Research and Innovation Fund (NSRF)
via the Program Management Unit for Human Resources \& Institutional
Development, Research and Innovation under Contract No. B16F640076;
Olle Engkvist Foundation under Contract No. 200-0605; STFC (United
Kingdom); Suranaree University of Technology (SUT), Thailand Science
Research and Innovation (TSRI), and National Science Research and
Innovation Fund (NSRF) under Contract No. 160355; The Royal Society,
UK under Contracts Nos. DH140054, DH160214; The Swedish Research
Council; U. S. Department of Energy under Contract No.
DE-FG02-05ER41374

\bibliography{ref}

\end{document}